\documentclass{aa}
\usepackage{graphicx}
\usepackage{txfonts}
\usepackage{longtable}
\usepackage[authoryear]{natbib}
\bibpunct{(}{)}{;}{a}{}{,}
\begin{document}
\newcommand{\kms}{km~s$^{-1}$}
\newcommand{\hcop}{HCO$^+$}
\newcommand{\nh}{NH$_3$}
\newcommand{\hi}{H\,{\small I}}
\newcommand{\chisq}{$\chi^2$}

%
%
  \title{A molecular cloud within the light echo of V838 Monocerotis}

  \author{
  T. Kami\'{n}ski\inst{1,2}
  \and R. Tylenda\inst{1}
  \and S. Deguchi\inst{3}
         } 


  \institute{Department for Astrophysics, N. Copernicus
            Astronomical Center, Rabia\'{n}ska 8,
            87-100 Toru\'{n}, Poland,  
  \email{tylenda@ncac.torun.pl}\label{inst1}
  \and Max-Planck Institut f\"ur Radioastronomie, 
       Auf dem H\"ugel 69, 53121 Bonn, Germany
  \email{kaminski@mpifr-bonn.mpg.de} \label{inst2}
  \and Nobeyama Radio Observatory, 
       National Astronomical Observatory of Japan,
       462-2 Nobeyama, Minamimaki, Minamisaku, Nagano 384-1305, 
  \email{deguchi@nro.nao.ac.jp}\label{inst3}}
	    

\abstract
 {V838 Mon is an eruptive variable, which exploded in 2002. It displayed the most spectacular light echo ever observed. However, neither the origin of the reflecting matter nor the nature of the 2002 outburst have been firmly constrained.} 
 {We investigate the nature of the CO radio emission detected in the field of the light echo. In particular, we explore its connection to the  echoing dust around V838 Mon.}
 {We observed the echo region in multiple CO rotational transitions. We present
 and analyse maps of the region obtained in the $^{12}$CO(1--0) and 
 (3--2) lines. In addition, deep spectra at several positions were acquired in $^{12}$CO(1--0), (2--1), (3--2), and $^{13}$CO(1--0), (2--1). Radiative transfer modelling of line intensities
 is performed for chosen positions to constrain the kinetic temperatures and densities. We derive global parameters (e.g. mass, distance, total column density) of the emitting cloud.} 
 {We found that a compact molecular cloud is located within the echo region. The molecular emission is physically connected to the dusty environment seen in the optical echo and they both belong to the same translucent cloud. The interstellar nature of the cloud is confirmed by its high mass of 90--150~M$_{\sun}$. We propose that the cloud consists of material remaining after the formation of the cluster to which V838 Mon belongs. This indicates that the eruptive star has young age (3--10~Myr). 
}{} 

\keywords{radio lines: ISM - ISM: clouds - ISM: molecules - 
          stars: individual: V838~Monocerotis} 
        
\titlerunning{CO emission in the light echo}
\authorrunning{Kami\'{n}ski}
\maketitle

\section{Introduction} \label{intro}
V838 Mon is an unusual variable, whose nova-like eruption was observed at the beginning of 2002 \citep{munari}. The object displayed a peculiar light curve with at least three distinct maxima seen in the optical. At maximum light, V838 Mon reached a luminosity of 10$^6$~L$_{\sun}$ and displayed the spectrum of an F-type star. The outburst lasted for about 3 months and was ended with a steep decline, e.g., a decline in $V$-band brightness of about 8~mag within one month. The decline was accompanied by a rapid decrease in the effective temperature \citep{tyl05} and the object eventually turned into a cool supergiant with a spectral type later than M10 \citep{evans}. V838 Mon has retained its low temperature since then and its spectral type is now close to M 6--7 \citep{keckI,uves}. The observational characteristics of the object during and after the outburst can be found in \citet{MunariASP}.

Several different explanations of the eruption of V838 Mon have been proposed, which were critically discussed by \citet{merger}. The most promising model appears to be that of a stellar collision followed by a merger \citep{soker,merger}. A crucial point in establishing the nature of the outburst is to constrain the evolutionary status of the progenitor of V838 Mon. 

V838 Mon displayed a spectacular light echo, which has been monitored since 2002 February. Multi-epoch images of the echo show a diffuse nebula surrounding the star that exhibits a complex morphology \citep{bond03,spirala}. Polarimetric observations of the echo were used to derive a geometric distance of 6~kpc \citep{sparks}. The nature of the dusty material responsible for the echo remains unclear. \citet[][see also \citealp{tylprog}]{tylecho} interpret it as an interstellar cloud, while \citet{bond2} advocates a circumstellar origin. If the echo material was ejected by the progenitor of V838 Mon, its properties (e.g. its mass ) should be consistent with the nature of the progenitor object. Constraining the nature of the echoing cloud is therefore important to more clearly understand the enigmatic eruption from 2002.  

Emission in the lowest CO rotational transitions was detected at the position of V838 Mon and in its close vicinity \citep{kami1,deguchi07}. First observations of the molecular line emission suggested that the source is extended. These observations were however of poor spatial resolution or incomplete sky coverage, so the connection of the molecular gas to either V838 Mon or the light-echo material was  unclear. \citet{kami2} observed 13 positions within the echo region with the IRAM 30-m telescope and found that the emission is indeed extended and shows a complex spatial distribution. However, its relation to the light echo material remained unclear. 

In this paper, we present comprehensive, follow-up observations of the echo region in several CO rotational transitions. The observations are described in Sect.~\ref{obsred} and the results are presented in Sect.~\ref{results}. The observations reveal the  presence of a molecular cloud within the echo region. We constrain its basic physical parameters in Sect.~\ref{analysis}. In Sect.~\ref{connection}, we show that the molecular cloud is physically related to the dusty medium seen in optical echo images. In Sect.~\ref{disc}, we discuss the nature of the cloud. 

\section{Observations and data reduction \label{obsred}}

A restricted region around V838~Mon was mapped in the
$^{12}$CO(1--0) and (3--2) transitions, for which the observations, data
reduction, and map assembly procedures are described in
Secs.~\ref{map10} and \ref{map32}. In addition, some positions within
the mapped region were observed in the standard on-off (position switching, PSW)
mode. These observations include integrations in the $^{12}$CO(1--0),
$^{13}$CO(1--0), and HCO$^{+}$(1--0) lines. Technical details about
these measurements are given in Sect.~\ref{onoff1}.

In this paper, we often make a reference to observations previously 
reported in \cite{kami2}. We summarize these data in Sect.~\ref{iram}. Some figures in this paper
also include an archival optical image of the light echo performed with the HST. The source of the image and the applied reduction procedure are shortly described in Sect.\ref{hst}.    

All velocities in this paper are given with respect to the local standard of rest
(LSR) and celestial coordinates are given for the epoch J2000.0.   

\subsection{BEARS map in $^{12}$CO(1--0)}\label{map10}

Mapping in $^{12}$CO $J$=1--0 (115.27\,GHz) was performed with the BEARS
multi-beam array \citep{bears1,bears2} installed on the 45-m telescope of the Nobeyama Radio
Observatory (NRO), Japan. The array consists of 25 detectors arranged in a
$5\times5$ grid separated by 41\farcs1 from each other. The half-power beam width (HPBW) of the 45-m telescope at
115\,GHz is 14\farcs9. The map was obtained in the on-the-fly (OTF)
technique \citep{otf} with  
$5''$ spacing between successive scan rows. The whole region of interest was
observed several times and successive coverages were obtained in orthogonal directions. The position angle of the 
array during observations was PA\,$=58\degr$ east of north. We
mapped the vicinity of V838~Mon in two steps. First, in 2008 April 2--4
we observed an area of $320''\times320''$ centerd at the position
$\alpha=07^h04^m05\fs70$, $\delta=-03\degr50'25\farcs0$. In this area,
only the inner part of $137''\times137''$ was fully 
sampled. Next, on 2008 April 12 we mapped a larger area of
$640''\times640''$ with full sampling within the inner region of $457''\times457''$. 

The central off position used for the OTF observations was at
$\alpha=07^h05^m25\fs70$, $\delta=-04\degr03'25\farcs0$. This position was found to be free of CO(1--0) emission
in the sensitive observations with the Delingha telescope reported in \cite{kami1}.  

All the BEARS data were acquired with the array of digital autocorrelation
spectrometers \citep{sorai} in a mode that provides a bandwidth of
32\,MHz ($83$\,\kms) per detector and a spectral resolution of 31.25\,kHz ($0.081$\,\kms). 
The system noise temperature, including atmospheric
contributions, was typically $T_{\rm sys}=400$\,K (DSB).     

Data were calibrated with the standard chopper-wheel method \citep{chopper}. The
calibration of the antenna temperature, $T_A^*$, was controlled by 
observations of a strong CO source in the Orion~KL region, located at
$\alpha=05^h35^m14\fs5$, $\delta=-05\degr22'30\farcs4$. Day-by-day measurements demonstrated that the calibration was stable to within 14\%
(3$\sigma$ noise). Pointing, crucial for high-frequency observations at the 45-m
telescope, was performed on strong SiO 
maser sources every $\sim$1~h. The observations were carried out in slow
wind (i.e. with an average wind speed lower than 5~m~s$^{-1}$) or no wind
conditions and the pointing should be accurate to within $\sim9''$ (3$\sigma$). 

Data reduction and map generation were performed with {\it
NOSTAR} 
\citep{otf}. Baselines of up to third order were subtracted from 
the spectra and the data were rescaled to single side-band (SSB) scale using conversion factors provided by NRO. The data were then converted to main beam brightness temperature units, $T_{\rm mb}=T_A^*/\eta_{\rm mb}$, using the main beam efficiency of $\eta_{mb}$=0.32. To reduce the scanning noise effect, the data were regridded and convolved into a map using
the basket-weave method \citep{otf2}. 
The data were convolved with a Gaussian-tapered first-order Bessel function
(Bessel$\times$Gauss) as the {\it gridding convolution function}. For a 
grid spacing of 7\farcs5 in the final map, the described procedures resulted in an effective
beam-size of 19\farcs2 (HPBW).       

\subsection{Single position measurements with BEARS}\label{onoff1}
Several positions within the echo region were observed in $^{13}$CO
$J$=1--0 (110.20\,GHz), \hcop\ $J$=1--0 (89.19\,GHz), and $^{12}$CO
$J$=1--0 in the standard  PSW mode. The
measurements were performed with BEARS. The HPBWs and
main-beam efficiencies for the NRO 45-m telescope at the observed
frequencies are given in Table~\ref{tabtel}. 

Measurements in $^{13}$CO and $^{12}$CO were obtained for two
central positions (note that the actual number of points observed
with BEARS is 25 for every specified central position). One of the
observed positions was the emission maximum found on the
$^{12}$CO(3--2) map (see Sect.~\ref{map32}), which we hereafter call P32. It is located at
$\alpha=07^h04^m05\fs40$, $\delta=-03\degr50'13\farcs0$. We
also observed the position with $^{12}$CO(1--0) maximum, as identified in the BEARS map. This position has coordinates $\alpha=07^h04^m05\fs52$, $\delta=-03\degr50'00\farcs0$, which we refer to hereafter as P10. In the case of the HCO$^+$(1--0) line, only P10 was observed. Additional observing details for the
PSW spectra are listed in Table~\ref{tabobs}.

The calibration and data reduction of the PSW observations was completed using analogous procedures as for the OTF observations with BEARS (see Sect.~\ref{map10}). Data reduction was performed using the {\it NEWSTAR} package \citep{newstar}. 

\begin{table}
\caption{NRO 45-m telescope specification.}
\label{tabtel}
\centering
\begin{tabular}{cccc}
\hline
transition&frequency&HPBW&$\eta_{\rm mb}$\\
          &  [GHz]  &[$''$]&\\
\hline
$^{12}$CO $J$=1--0&115.271204&14.9&0.32\\
$^{13}$CO $J$=1--0&110.201353&14.9&0.40\\
  HCO$^+$  $J$=1--0&~89.188518&18.2&0.44\\
\hline
\end{tabular}
\end{table}
\begin{table*}\begin{minipage}[t]{\hsize}
\centering
\caption{Characteristics of the PSW observations with BEARS.}
\label{tabobs}
\renewcommand{\footnoterule}{}  
\begin{tabular}{ccccccrc}
\hline
line&position&date&PA\footnote{Position angle of BEARS}&spec. res.&$T_{\rm
  sys}$\footnote{DSB values}&\multicolumn{1}{c}{$t_{\rm ON}$}&rms\footnote{in the
  $T_{\rm mb}$ (SSB) scale, per the specified resolution element}\\ 
          &        &    &[\degr]& [\kms]   &       [K]        &            &[K]\\
\hline
$^{12}$CO(1--0)&P32\footnote{$\alpha=07^h04^m05\fs40$, $\delta=-03\degr50'13\farcs0$}
                      &2008.04.02&13&0.081&406& 1m20s&1.78\\
                &P10\footnote{$\alpha=07^h04^m05\fs52$, $\delta=-03\degr50'00\farcs0$}
                      &2008.04.15&58&0.081&390&14m20s&0.59\\
\hline
$^{13}$CO(1--0)&P32$^d$&2008.04.05/06&13&0.085&360&1h40m&0.09\\
                  &P10$^e$&2008.04.09   &13&0.085&390&1h10m&0.13\\
\hline
 HCO$^+$(1--0)&P10$^e$&2008.04.14&45&0.105&240&  54m&0.09\\
\hline
\end{tabular}\end{minipage}
\end{table*}
%
%
\subsection{Map in $^{12}$CO(3--2)}\label{map32}
The map in $^{12}$CO $J$=3--2 (345.7956\,GHz) was obtained with
HARP (Heterodyne Array Receiver Program for the B-band) \citep{harp1,harp2}
newly installed on the James Clerk Maxwell Telescope (JCMT). The
observations were performed during several periods between 2007
December 27 and 2008 January 17, with the majority of the data (60\% of total
integration time) being obtained on 2008 January 17. HARP is an array of 16
detectors, only 14 of which were operational at the time of our
observations. The detectors are arranged in a $4\times4$ grid,
separated by $30''$. The beam-size of each detector is $14''$ (HPBW). The map was
obtained in the {\it jiggle-position switch} mode with the {\it HARP5 jiggle}
pattern, which provides measurements on a $6''\times6''$ 
grid. Because of the non-operational detectors and to fully cover the  
interesting area, observations were carried out with different array
orientations, mainly with position angles of PA\,$=45\degr$ and
PA\,$=0\degr$. In addition, the map center was changed for different observing
runs. In consequence, the shape of the resultant map is irregular and in terms of sensitivity the map is rather inhomogeneous.   

The central sky reference (off) position was the same as in the NRO
observations, i.e. at $\alpha=07^h05^m25\fs70$,
$\delta=-04\degr03'25\farcs0$. Each jiggle position had
a separate off position. Observations were performed
in good weather conditions with a typical atmospheric opacity at 220\,GHz of $\tau_{\rm CSO}=0.10$ (as
measured by the tau-meter at the nearby Caltech Submillimeter
Observatory). The receiver noise temperature (no atmospheric
contribution) was in the range 109--201\,K for different detectors. Pointing was performed at 345\,GHz
regularly every 1\,h on strong  
stellar sources and the pointing accuracy should be within $4\farcs5$
(3$\sigma$). The calibration was controlled by observations of
standard targets, e.g. CRL618 and IRC+10216, and its stability was better
than about 10\%.      

As a back-end, we used the ACSIS autocorrelator in the configuration, which provides a 250\,MHz ($217$\,\kms) bandwidth and  a 31\,kHz ($0.027$\,\kms) channel resolution. 
The data were reduced with Starlink software, in
particular with the Kernel Application Package ({\it KAPPA}). Baselines of up to
third order were subtracted from all spectra and the data were regridded onto a
data-cube with a 6\arcsec$\times$6\arcsec\ spatial sampling.
The resultant data were then converted to a main beam brightness 
temperature with main beam efficiency of $\eta_{\rm mb}=0.63$.  

\begin{figure}
\centering
\includegraphics[angle=270, width=\hsize]{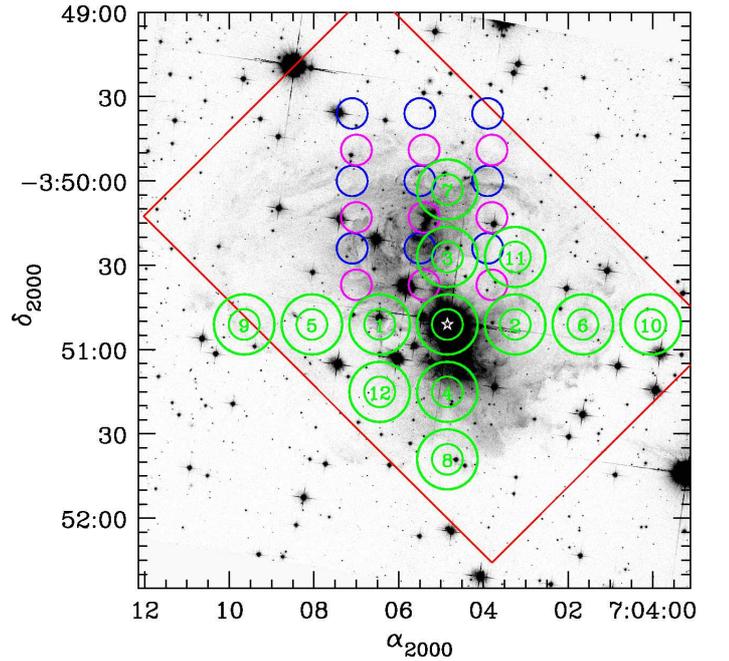}  
\caption{Selected regions observed in CO lines are shown on top of the optical image of the light echo. The red frame shows the main region covered by the CO(3--2) map obtained with HARP/JCMT in 2008 January. The green circles represent positions observed with the single-pixel receivers on the IRAM 30-m telescope in 2006 September. These positions are labelled with numbers introduced in \citet{kami2}. The sizes of the circles correspond to beam sizes (HPBW) at the observed frequencies, i.e. the larger ones  represent the CO(1--0) data, the smaller ones represent the CO(2--1) data.  The blue and magenta circles correspond to HERA observation obtained with the IRAM 30-m telescope in the CO(2--1) line in Jan. 2009.  The blue and magenta colours correspond to observations with the array centerd at positions P10 and P32, respectively. The position of V838 Mon is indicated with a star. The background image was obtained with HST/ACS in the F814W filter on 2006 September 10.} \label{iram_rys}
\end{figure}

\subsection{IRAM observations in the $^{12}$CO(1--0) and (2--1) lines}\label{iram} 
For our analysis and discussion, it is important for us to consider the data reported in
\cite{kami2}. These observations were obtained in 2006 September 27--28 with
the IRAM 30-m telescope using single-pixel receivers in the frequency switching mode. The data consist
of very sensitive measurements in the $^{12}$CO(1--0) and (2--1) lines for 13
positions within the echo region, as shown in Fig.~\ref{iram_rys}. More technical details about the
spectra can be found in \cite{kami2}.     

\subsection{HERA observations obtained in the $J$=2--1 line of $^{12}$CO and $^{13}$CO}
The last spectroscopic observations were performed on 2009 January 29 with the IRAM 30-m telescope in the $^{12}$CO(2--1) (230.54~GHz) and $^{13}$CO(2--1) (220.40~GHz) lines. The HERA (HEterodyne Receiver Array) \citep{hera} was used in the stare mode. This array consists of nine detectors arranged in a 3$\times$3 grid with a separation of 24\arcsec\ between neighbouring detectors. In our observations, the central detector was centerd on either P10 or P32 with the position angle of the array being set to zero. The observing mode was PSW with an off position located 600\arcsec\ east from the observed on position. The integration times, system temperatures, and the resulting rms values (in units of $T_{\rm mb}$) are given in Table~\ref{hera_tab}.

The VESPA autocorrelator was used as a back-end. The total bandwidth was 71~MHz (95~\kms) with a spectral resolution of 78~kHz (0.1~\kms). The beam-width at both observed frequencies is practically the same with HPBW=11\arcsec. Data were calibrated with the standard chopper-wheel method. The pointing was controlled by regular observations of strong continuum sources and should be of higher accuracy than 4\farcs5 (3$\sigma$). Some observations were obtained at very low source elevation ($\sim$18\degr).

The data were reduced using the {\it CLASS} package, which is part of the {\it GILDAS} software collection. Baselines were corrected by subtracting low-order polynomials from individual scans. For each position, the scans were averaged  and transformed to the $T_{\rm mb}$ scale using efficiencies 
from \citet{iramnewsletter}, i.e., $\eta_{\rm mb}$(220~GHz)=0.59 and $\eta_{\rm mb}$(230~GHz)=0.57. 

\begin{table}\begin{minipage}[t]{\hsize}
\centering
\caption{Characteristics of the HERA observations from 29 Jan. 2009.}\label{hera_tab}
\renewcommand{\footnoterule}{}  
\begin{tabular}{ccccc} \hline
line&position&$T_{\rm sys}$\footnote{SSB value}&$t_{\rm ON}$&rms\footnote{An average value for all of the 9 beams per resolution element}\\ 
 & & [K] & [min] & [mK]\\
\hline
$^{12}$CO $J$=2--1&P32&580& 44&118\\
                                  &P10&635& 44&133\\
\hline
$^{13}$CO $J$=2--1&P32&390&210&34\\
                                  &P10&400&210&35\\
\hline
\end{tabular}\end{minipage}
\end{table}

\subsection{HST optical observations of the light echo}\label{hst}
Multi-epoch observations of the light echo have been performed with HST on several occasions since 2002 and were partially published in \citet{bond03} and \citet{bond2}. For the sake of data presentation and the analysis, we used the data obtained on 2006 Sept. 10 with the Advanced Camera for Surveys (ACS). The CCD frames were downloaded  from the Multimission Archive at the Space Telescope Science Institute (MAST). They were reduced with the standard pipeline and reprocessed with the {\it multidrizzle} procedure implemented in the {\it STSDAS} package of pyRAF.
  
\section{Results of the observations}\label{results}


\subsection{Maps of the $^{12}$CO(1--0) line}
\begin{figure*}
    \resizebox{\hsize}{!}{\includegraphics[angle=270]{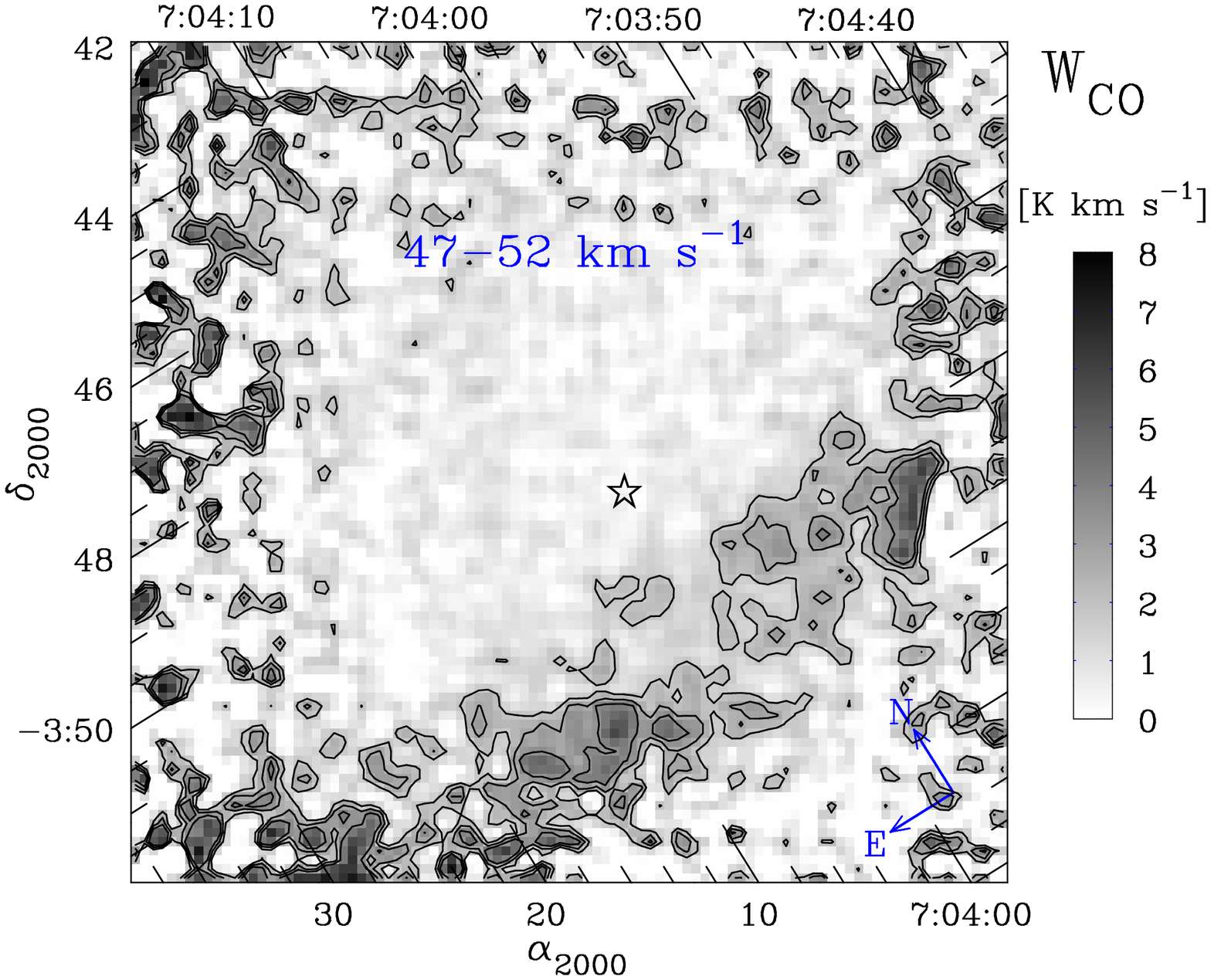}
                          \includegraphics[angle=270]{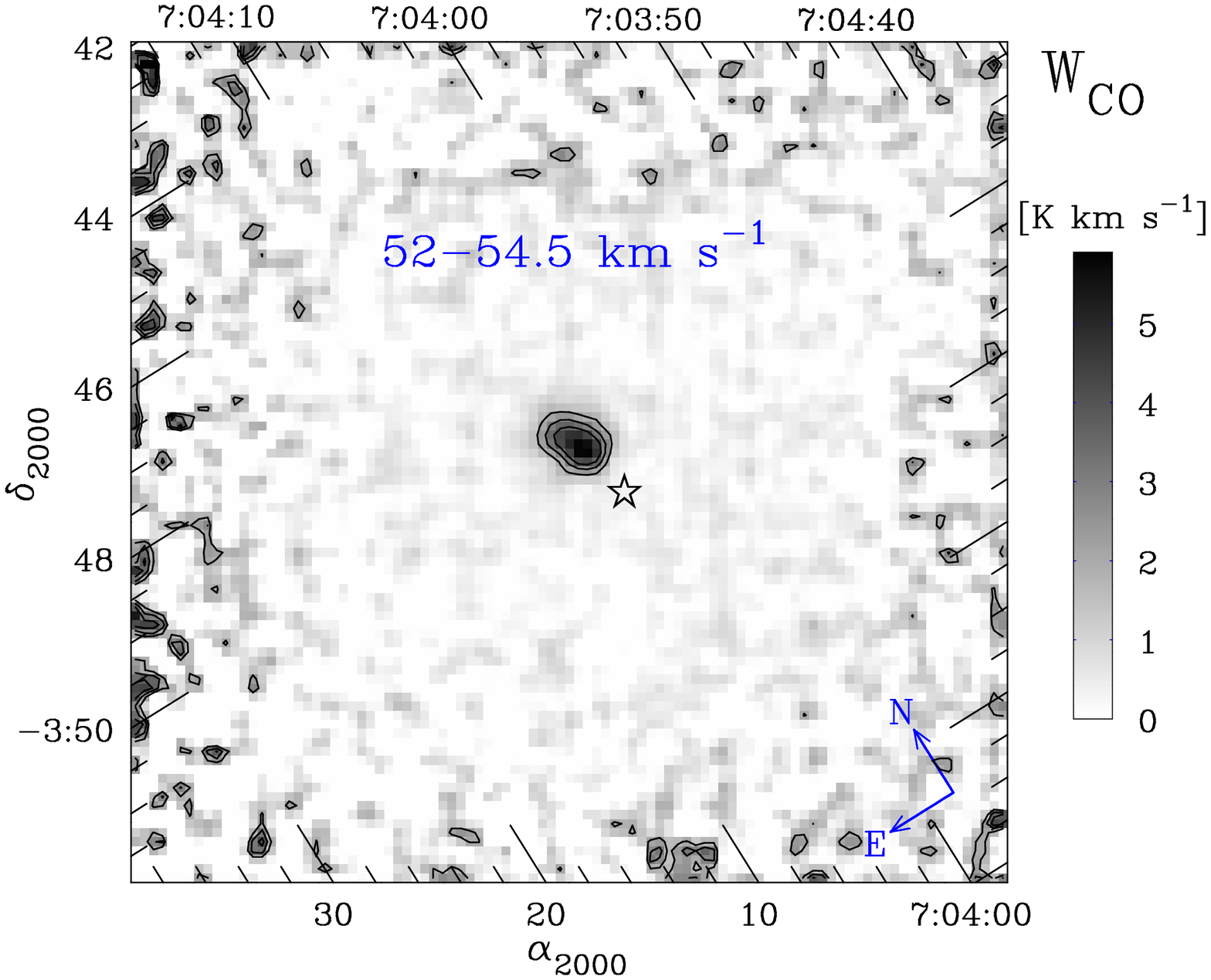}}
    \resizebox{\hsize}{!}{\includegraphics[angle=270]{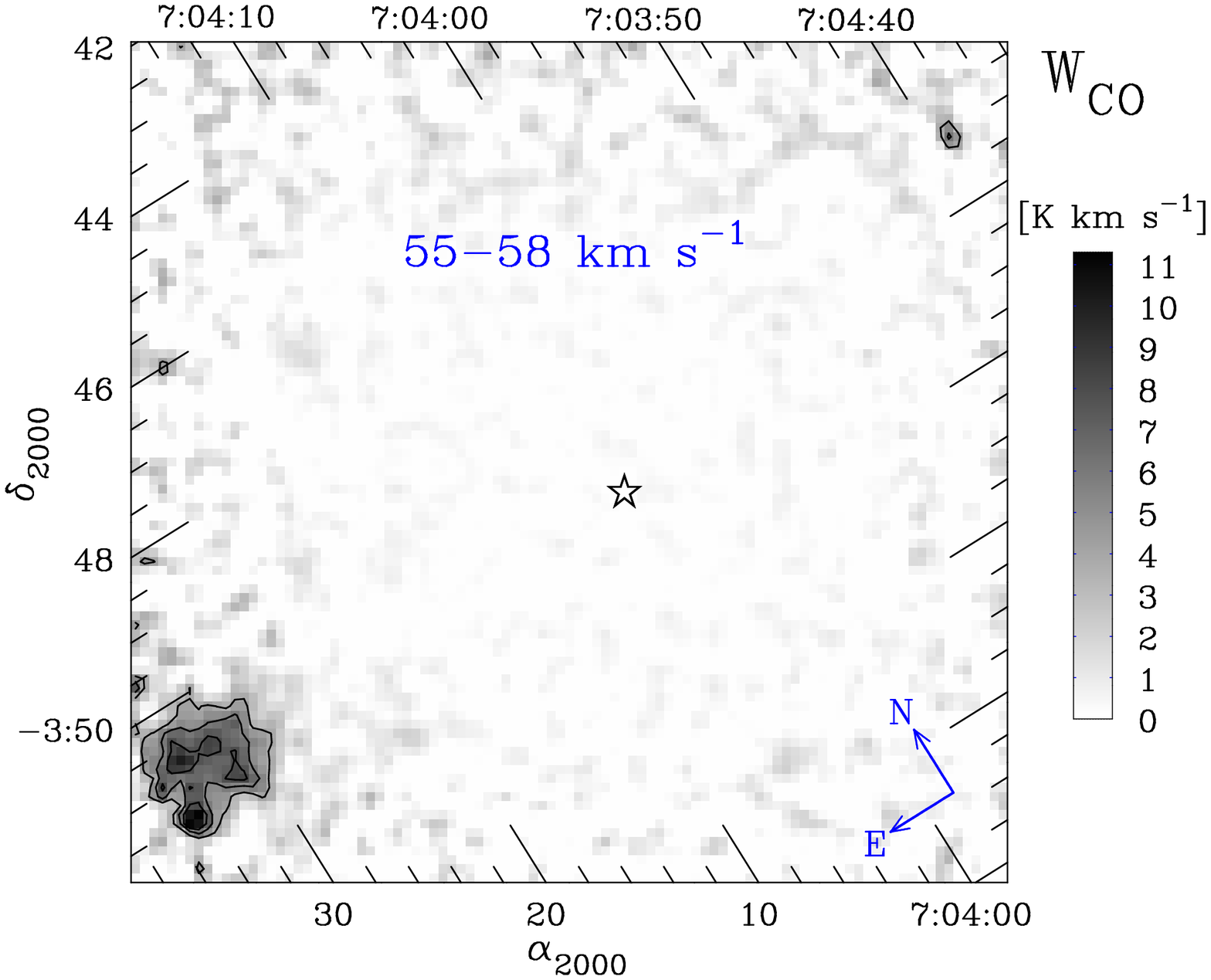}
                          \includegraphics[angle=270]{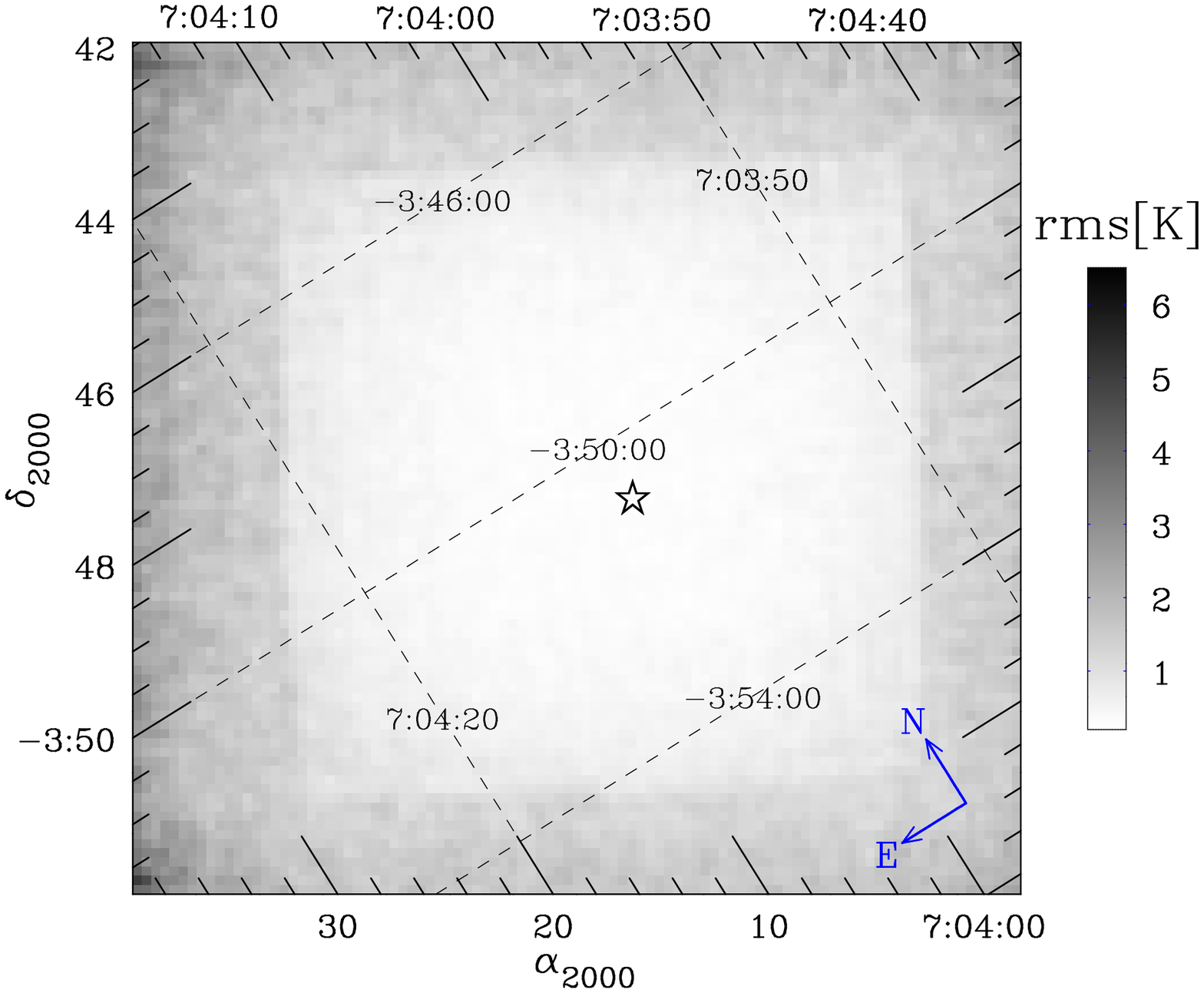}}
   \caption{{\it Upper left and right and lower left:} Integrated intensity maps in the $^{12}$CO(1--0) line obtained with the BEARS instrument at the NRO 45-m telescope. The signal was integrated in the velocity range indicated on each map. Contours are drawn for 30, 45, and 60\% of the maximal integrated signal. {\it Lower right:} The noise-level distribution in the $^{12}$CO(1--0) data cube (rms values are given per 0.1\,\kms\ bin). Dashed lines are drawn for constant declination and right ascension. Note that the frames are not aligned with the axis of the equatorial system. The star marks the position of V838 Mon.}\label{10maps}
\end{figure*}

A careful analysis of the $^{12}$CO(1--0) data-cube uncovers three kinematically distinct cloud complexes
in the mapped region. They can be assigned to the velocity ranges 47.0--52.0~\kms, 52.0--54.5~\kms, and 55.0--58~\kms. Integrated
intensity maps for these velocity intervals, together with 
a map of the rms noise distribution of the cube, are shown in Fig.~\ref{10maps}. The maps were
obtained with a $7\farcs5\times7\farcs5$ grid spacing. 

The clouds that appear in the velocity range 47.0--52.0~\kms\ are
located in a declination strip of (approximately) 
$-3\degr51'13''\leq\delta\leq-3\degr55'30''$. 
The strip is seen along the whole map 
and forms an elongated complex, which probably extends outside the
mapped region. The $^{12}$CO(1--0) emission in this molecular region is weak,
and has multiple local maxima centers with peak temperatures reaching
$\sim$3~K. This molecular complex is very diffuse and consists of
numerous patchy clouds.   

In the velocity range 52.0--54.5~\kms, only one cloud can be seen. It is located north with respect to V838~Mon. Although certainly extended, the cloud has a compact structure. It resembles a comet with a core close to V838~Mon, and has a length of 93\arcsec\ (as measured for an isophot at 10\% of the intensity peak) along its longer axis. It is approximately aligned in the north-south direction
(see Fig.~\ref{echoall}). The integrated intensity peak is located at $\alpha=07^h04^m05\fs52$, 
$\delta=-03\degr50'00\farcs00$. This position is 52\arcsec\ away from V838~Mon. At a majority of positions, the line shapes are Gaussian and have no structure, although  
some lines are slightly asymmetric. The typical full width at half maximum (FWHM) is $\Delta V=0.9$~\kms. 

In the under-sampled part of our map, at approximately
$\alpha=07^h04^m33^s$, $\delta=-03\degr51'00''$, an emission
region can be found at velocities 55.0--58.0~\kms. This cloud
is seen in the very noisy part of the map, close to the map's edge,
where it is difficult to characterize the feature in more detail. 
%
\subsection{The HARP map in the $^{12}$CO(3--2) line}
\begin{figure*}
    \resizebox{\hsize}{!}{\includegraphics[angle=270]{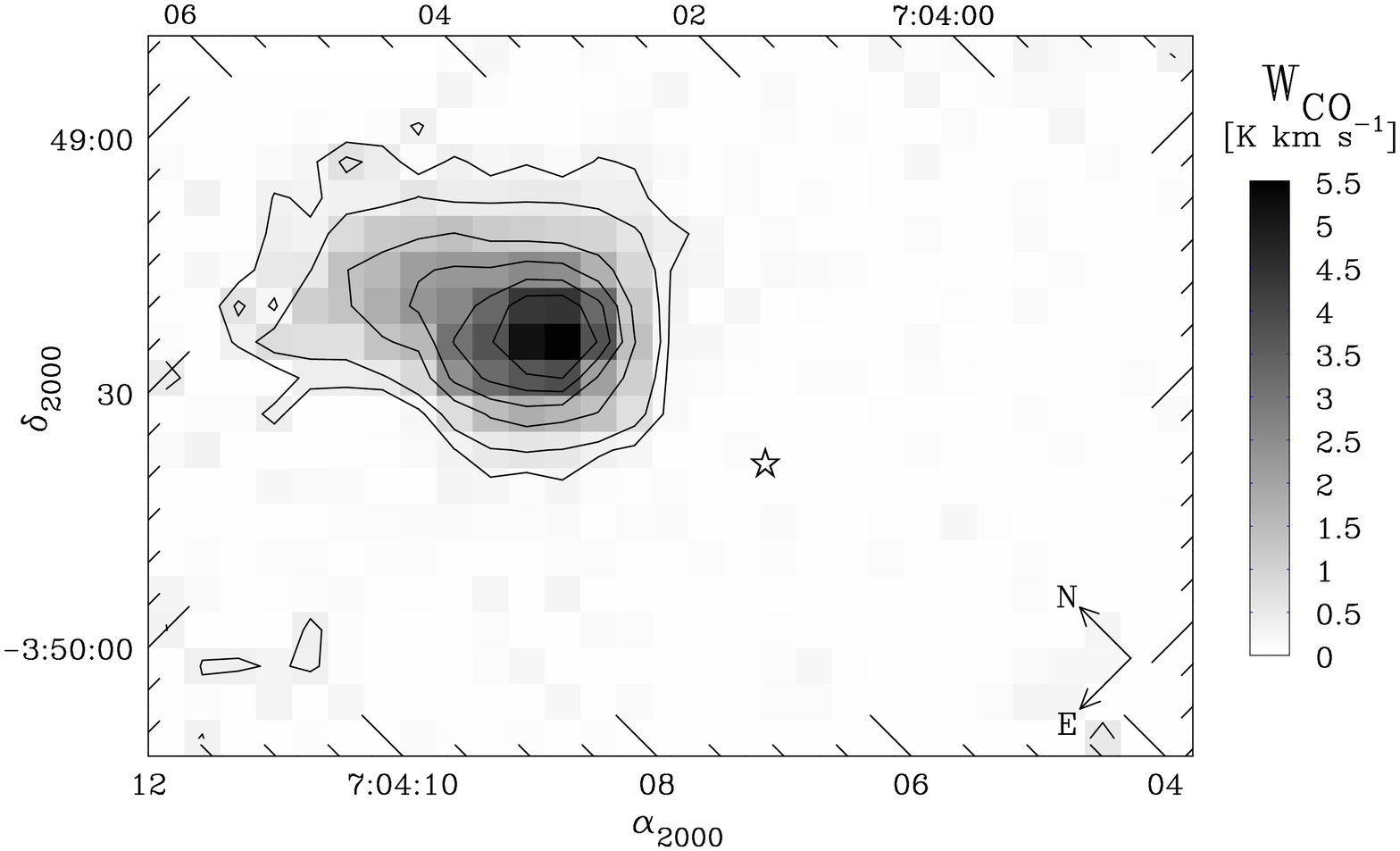} 
                          \includegraphics[angle=270]{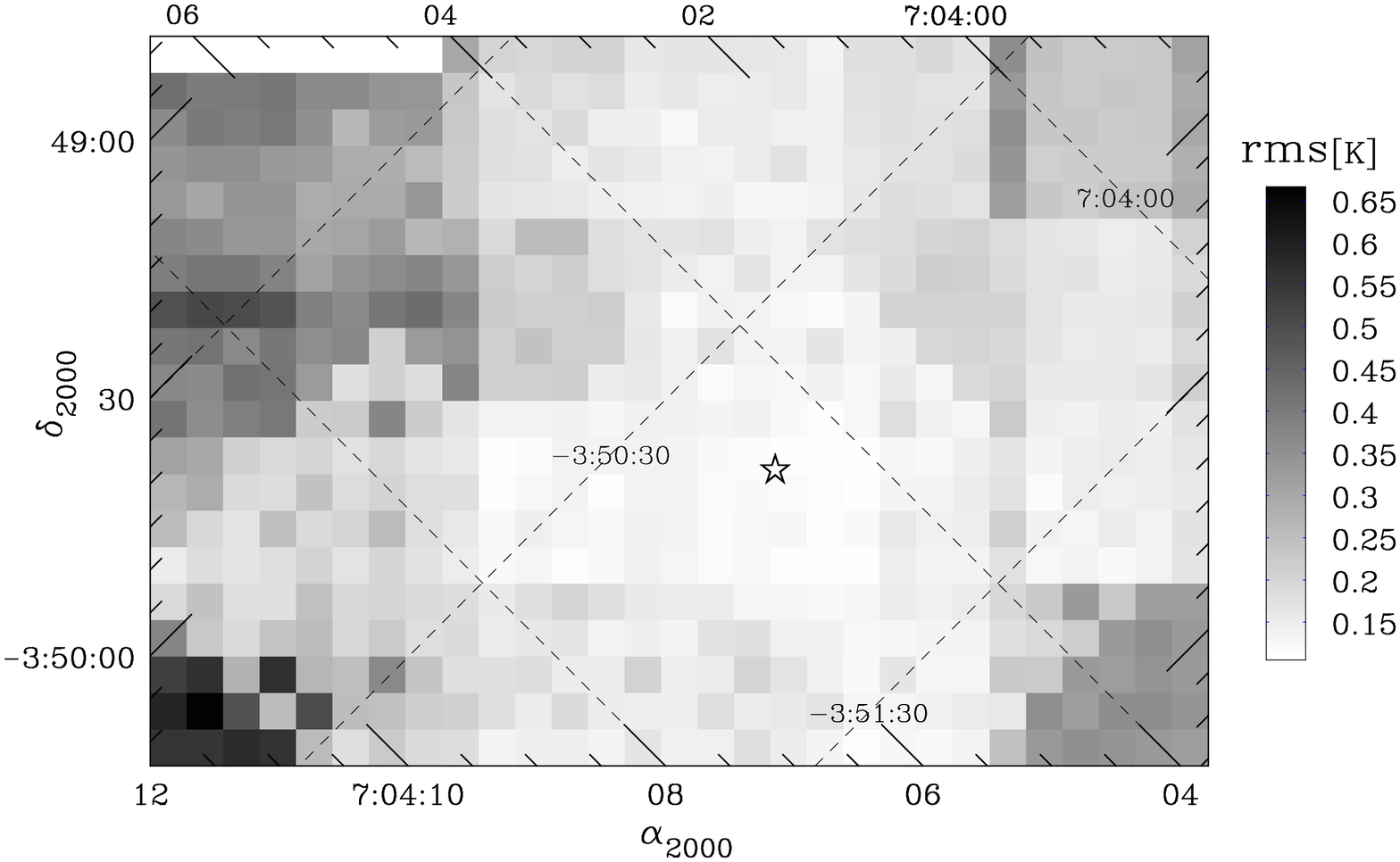}}  
   \caption{{\it Left:} Integrated intensity map of the CO(3--2) emission in the range 52--54.5~km~s$^{-1}$, as observed with HARP/JCMT. Contours are at 5, 10, 25, 40, 55, and 70\% of the peak emission. {\it Right:} The corresponding rms map, where rms is given per 0.1~km~s$^{-1}$ bin. Dashed lines are drawn for constant declination and right ascension. Note that the frames are not aligned with the axis of the equatorial system.}
\label{32maps}
\end{figure*}

The region observed with HARP is much less extended than the map obtained in CO(1--0) and corresponds to a central portion of the large OTF map. The HARP observations in the $^{12}$CO(3--2) line reveal an extended region of
emission in the velocity range 52--54.5\,\kms\ and in
approximately the same sky region as the cloud found in CO(1--0) data
at the same velocity range. A map of
integrated intensity and a
corresponding map of rms noise are presented in Fig.~\ref{32maps}
(see also Fig.~\ref{echoall}). The lines in the emission region are very narrow with a typical FWHM of $\Delta V = 0.9$\,\kms.

The general appearance of the cloud is similar to that of the
analogous cloud found in the CO(1--0) map. The feature seen in CO(3--2)
clearly has a cometary-like structure with a core oriented
towards the position of the star. The 10\% contour of integrated intensity has a dimension of $\sim$60\arcsec\ along the longer axis, and $\sim$40\arcsec\ in the orthogonal direction. The point with the strongest emission is located at $\alpha=07^h04^m05\fs4$, $\delta=-03\degr50'13\farcs0$, that is $39''$ from the position of V838 Mon. 


%
%
\subsection{Measurements of single positions observed with BEARS}

\subsubsection{$^{12}$CO(1--0)}
The observations in the $^{12}$CO(1--0) line at P10 and P32 as central positions were very short and intended to be control observations. The line was clearly detected only at the two central positions. The line parameters are listed in Table~\ref{allp10p32}. The measured intensities of these lines are important for understanding the uncertainties in the OTF data (see Sect.~\ref{othvsonoff}). 

\subsubsection{$^{13}$CO(1--0)}
The BEARS observations in the $^{13}$CO(1--0) line only resulted in detections for the central detector, i.e. at P10 and P32. These detections are at the level of 5$\sigma$. The line parameters are listed in Table~\ref{allp10p32}. The central velocities of the observed emission lines agree very well with the corresponding velocities for $^{12}$CO at the same positions. The $^{13}$CO lines seem to be narrower than their $^{12}$CO counterparts, but this is uncertain for these noisy spectra.

\subsubsection{\hcop\ (1--0)} \label{hcop} 
In the spectrum acquired in the \hcop (1--0) transition, no line is detected and only an upper limit to  a potential detection can be given. For a 0.105\,\kms\
channel bin, the limit on the peak intensity is $T_{\rm mb}$=0.27~K (3$\sigma$). This value is only half of the peak intensity of the $^{13}$CO (1--0) line at the same position. 

\begin{table}
\begin{minipage}[t]{\hsize}
\caption{Characteristics of the $J$=1--0 lines of $^{12}$CO and $^{13}$CO observed at positions P10 and P32 with the central BEARS receptor in Apr. 2008. } 
\label{allp10p32} \centering
\renewcommand{\footnoterule}{}  
\begin{tabular}{c|ccccc}
\hline
position&V$_{\rm LSR}$&$\Delta$V&T$_{\rm mb}^{\rm max}$&$I_{\rm CO}$&rms\\
&[km\,s$^{-1}$]&[km\,s$^{-1}$]&[K]&[K\,km\,s$^{-1}$]&[mK]\\[2pt]    
\hline 
\multicolumn{1}{c}{}&\multicolumn{5}{|c}{$^{12}$CO(1$-\,$0) (OTF)}\\
\hline 
P10&53.2& 0.9& 6.10$^a$& 5.59&289.9\footnote{per 0.081 \kms\ channel}\\
P32&53.3& 0.9& 6.46$^a$& 6.33&287.4$^a$\\ 
\hline 
\multicolumn{1}{c}{}&\multicolumn{5}{|c}{$^{12}$CO(1$-\,$0) (PSW)}\\
\hline 
P10&53.4& 0.8& 8.13$^a$& 6.71&590.0$^a$\\
P32&53.5& 0.9& 8.07$^a$& 6.77&1782$^a$\\ 
\hline 
\multicolumn{1}{c}{}&\multicolumn{5}{|c}{$^{13}$CO (1$-\,$0) (PSW)}\\
\hline 
P10&53.3&0.7&0.59$^b$&0.45&129.0\footnote{per 0.085 \kms\ channel}\\
P32&53.1&0.5&0.51$^b$&0.27&~93.8$^b$\\
\hline 
\end{tabular}
\end{minipage}
\end{table}

\subsection{Measurements for single positions observed with HERA}
\begin{figure*}
 \includegraphics[angle=270, scale=0.5]{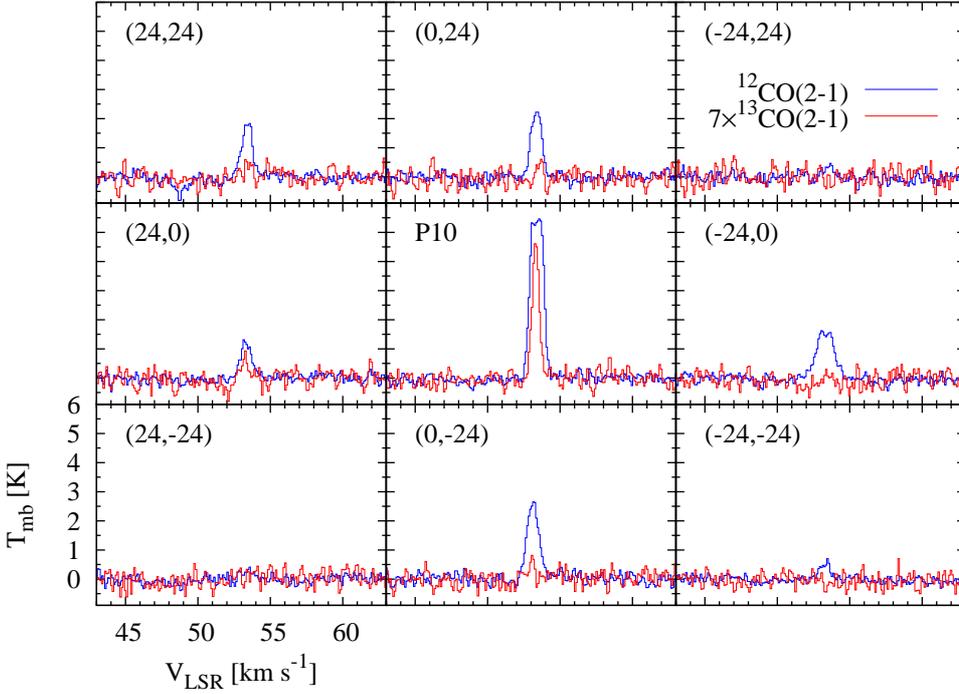} 
 \caption{Spectra acquired using the HERA array in Jan. 2009 with P10 as the central position. Spectra obtained in $^{12}$CO(2--1) are presented with a blue line, and those for $^{13}$CO(2--1) are shown with red line. The offsets of the observed positions are indicated in the left upper corner of each panel and are given with respect to P10.}   \label{hera_spec10}
\end{figure*}

\begin{figure*} \includegraphics[angle=270, scale=0.5]{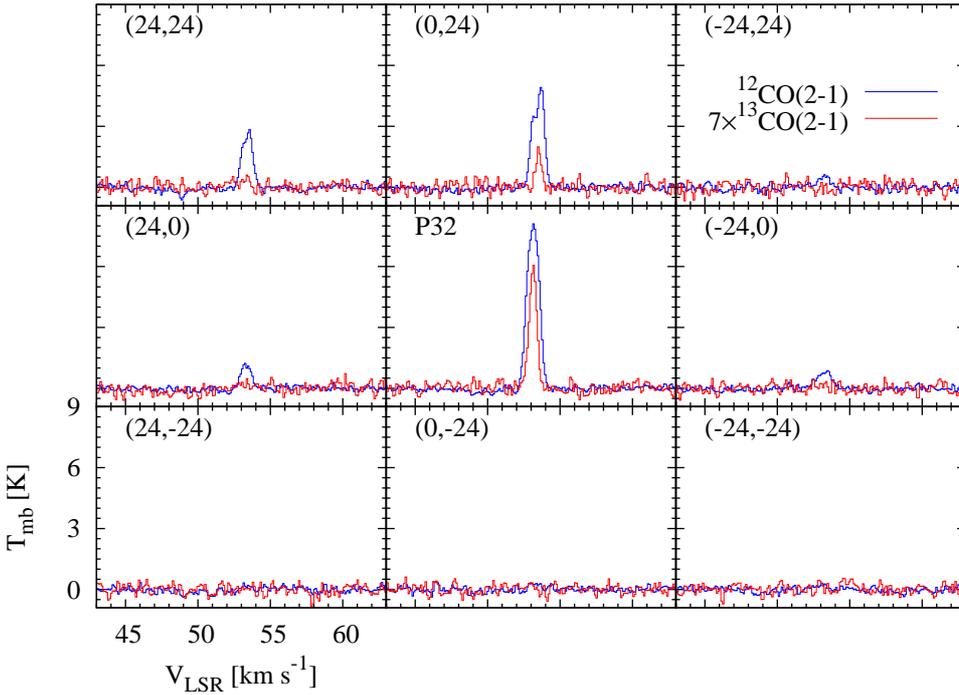}
 \caption{The same as in Fig.~\ref{hera_spec10}, but where P32 is the central position. The offsets are given with respect to P32.} 
 \label{hera_spec32}\end{figure*}

Spectra acquired with the HERA array in 2009 January are shown in Figs.~\ref{hera_spec10} and \ref{hera_spec32}. Line measurements are shown in Tables \ref{hera_lines10} and \ref{hera_lines32}. Among the positions observed with HERA centerd on P10, no emission in $^{12}$CO(2--1) was detected at the offset (24,--24). At positions (--24,--24) and (--24,24), very weak emission is seen with an integrated intensity below 1~K~\kms. At the remaining six positions, the lines are strong, with the strongest line observed at P10. The line profiles analysed at the original resolution are irregular. In at least two cases, i.e. at P10 and (--24,0), the profile consists of two components forming a tight blend. For two positions with strong $^{12}$CO emission, the $^{13}$CO line was also detected. While at P10 the line is strong, the line at (24,0) is detected at the level of 3$\sigma$. The emission lines of $^{13}$CO are clearly narrower than the corresponding lines of  $^{12}$CO.

In the case of observations centerd at P32, the $^{12}$CO(2--1) line was detected only at the six northern positions, i.e. no lines are seen at (24,--24), (0,--24), and (--24,--24). The observed line profiles are again irregular, but the line is clearly double only at (0,24). The isotopologue line  $^{13}$CO(2--1) is only detected at P32 and (0,24). These lines are almost half as narrow as their $^{12}$CO counterparts. 

\begin{table}\begin{minipage}[t]{\hsize}
\caption{Measurements for lines observed with HERA/IRAM in Jan. 2009 with P10 as the central position. The values of T$_{\rm mb}^{\rm max}$ are for channels of 0.102~\kms ($^{12}$CO) and 0.106~\kms ($^{13}$CO). Offsets are given with respect to P10.}\label{hera_lines10}
\centering \renewcommand{\footnoterule}{}  
\begin{tabular}{rr|cccc}
\hline
\multicolumn{2}{c|}{Offset wrt P10}&V$_{\rm LSR}$&$\Delta$V&T$_{\rm mb}^{\rm max}$&$I_{\rm CO}$\\ 
\multicolumn{2}{c|}{($\Delta\alpha$\arcsec,$\Delta\delta$\arcsec)}&[km\,s$^{-1}$]&[km\,s$^{-1}$]&[K]&[K\,km\,s$^{-1}$]\\    
\hline 
\multicolumn{2}{c|}{}&\multicolumn{4}{c}{$^{12}$CO(2--1)}\\ 
\hline
--24&--24& 53.3& 0.9& 0.52&  0.49\\ 
--24&   0& 53.3& 1.3& 1.63&  2.33\\ 
--24&  24& 53.5& 1.0& 0.47&  0.51\\ 
   0&--24& 53.1& 1.0& 2.68&  2.77\\ 
   0&   0& 53.4&1.0&~6.10\footnote{The line profile is not Gaussian. The peak of the profile is at T$_{\rm mb}^{\rm max}$=5.47~K and the signal summed over the profile gives $I_{\rm CO}$=6.15~K\,km\,s$^{-1}$}& ~6.46$^a$\\ 
   0&  24& 53.4& 0.8& 2.31&  2.01\\ 
  24&   0& 53.3& 0.9& 1.23&  1.16\\ 	
  24&  24& 53.4& 0.8& 1.87&  1.68\\ 
\hline
\multicolumn{2}{c|}{}&\multicolumn{4}{|c}{$^{13}$CO(2--1)}\\ 
\hline
  0&    0&53.3&0.6&0.67&0.42\\
 24&   0&53.2&0.6&0.10&0.06\\ 
\hline
\end{tabular} \end{minipage} \end{table}

\begin{table}\begin{minipage}[t]{\hsize}
\caption{The same as in Table~\ref{hera_lines10} but where P32 is the central position.}\label{hera_lines32}
\centering \renewcommand{\footnoterule}{}  
\begin{tabular}{rr|cccc}
\hline
\multicolumn{2}{c|}{}&\multicolumn{4}{c}{$^{12}$CO(2--1)}\\ 
\multicolumn{2}{c|}{Offset wrt P32}&V$_{\rm LSR}$&$\Delta$V&T$_{\rm mb}^{\rm max}$&$I_{\rm CO}$\\ 
\multicolumn{2}{c|}{($\Delta\alpha$\arcsec,$\Delta\delta$\arcsec)}&[km\,s$^{-1}$]&[km\,s$^{-1}$]&[K]&[K\,km\,s$^{-1}$]\\    
\hline
--24&   0& 53.3& 1.2& 0.81& 1.04 \\ 
--24&  24& 53.3& 0.8& 0.55& 0.48 \\ 
   0&   0& 53.1& 1.0& 8.38& 8.54 \\ 
   0&  24& 53.5& 1.0& 4.67& 4.74 \\  
  24&   0& 53.3& 0.8& 1.21& 1.02 \\  
  24&  24& 53.4& 0.8& 2.76& 2.47 \\  
\hline
\multicolumn{2}{c|}{}&\multicolumn{4}{|c}{$^{13}$CO(2--1)}\\ 
\hline
    0&   0&53.1&0.6&0.85&0.56\\
    0& 24&53.5&0.5&0.26&0.13\\
\hline
\end{tabular} \end{minipage} \end{table}

\subsubsection{Comparison between different data sets for the $^{12}$CO(1--0) line and calibration issues}\label{othvsonoff}

To help us understand the quality of the data discussed in this paper, we compared spectra obtained with different telescopes and observing modes. The comparison is made for the $^{12}$CO(1--0) line, which was observed at a few positions over several observing runs. The positions P10 and P32 were observed with the BEARS array in the OTF and PSW modes. Two positions with strong lines at offsets (--8,--14) and (--8,10) with respect to P32 (named respectively Off3 and Off7 in \citet{kami2}) were observed in the frequency-switching mode with single pixel receivers at the IRAM 30-m telescope,  as well as with the NRO 45-m telescope in the OTF mode. Spectra for these four positions were extracted from the gridded OTF data cube. Differences in coordinates of the extracted OTF spectra with respect to those defining the four positions should be much smaller than the formal pointing errors of the OTF observations. 

We first compare the data obtained with BEARS in the PSW and OTF modes. The emission lines are on average 1.4 times weaker in the OTF data than in the PSW observations. This difference exceeds the formal calibration error expected for our BEARS observations ($\sim$14\%). Some velocity shift between the two data sets can also be seen. 

When the OTF spectra are compared with the IRAM data, the discrepancy is even more pronounced. The OTF line intensities are 1.6--1.9 weaker than in the IRAM data.  Moreover, the line profiles for the Off3 position are significantly different in the two spectra. Although many factors may be responsible for the noticeable differences, the problematic pointing of the 45-m telescope may be the main source of errors. This problem may be applicable to both OTF and PSW observations and all the BEARS results should be treated with special care as it is probable that they are less accurate than the formal uncertainties indicate.    

\subsection{$^{12}$CO(3--2) spectra extracted for selected positions}

In the analysis of line intensities at positions P10 and P32, we extracted relevant spectra from the HARP/JCMT data cube. The differences between the coordinates of the observed positions and those defining P10 and P32 are smaller than the pointing errors at JCMT. The results of line measurements for these spectra are given in Table~\ref{tab32}.  

\begin{table}\begin{minipage}[t]{\hsize}
\caption{Characteristics of the $^{12}$CO(3--2) emission lines in spectra extracted from the data cube obtained with HARP/JCMT. The values of rms and T$_{\rm mb}^{\rm max}$ are given for a bin of 0.1~\kms.} 
\label{tab32}
\centering
\renewcommand{\footnoterule}{}  
\begin{tabular}{cccccc}
\hline
position&V$_{\rm LSR}$&$\Delta$V&T$_{\rm mb}^{\rm max}$&$I_{\rm CO}$&rms\\
&[km\,s$^{-1}$]&[km\,s$^{-1}$]&[K]&[K\,km\,s$^{-1}$]&[mK]\\    
\hline 
P10&53.4&0.9&3.58&3.44&217\\
P32&53.2&1.0&5.29&5.60&219\\   
\hline
\end{tabular}
\end{minipage}
\end{table}

\section{Analysis: physical parameters of the molecular cloud}\label{analysis}
\subsection{Distribution}
\begin{figure}
 \includegraphics[height=\hsize,angle=270]{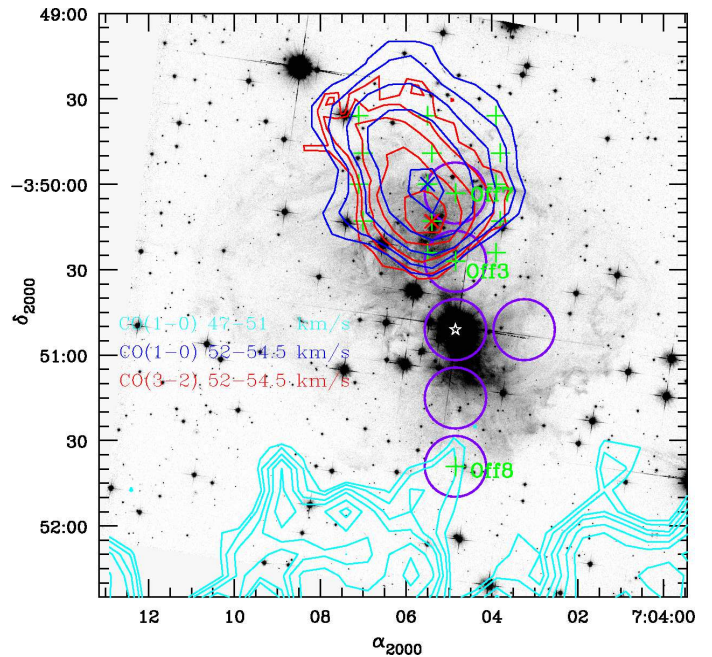}  
 \caption{Contours of $^{12}$CO emission overplotted on the
 light echo image from HST (obtained in Sept. 2006, filter F814W). The blue contours
 show the $^{12}$CO(1--0) emission integrated in the velocity range of
 52--54.5\,\kms\ (0.99, 1.35, 2.01, 3.26, and 5.66\,K~\kms), while
 cyan contours correspond to $^{12}$CO emission integrated in the range
 47--51~\kms\ (1.19, 1.34, 1.57, 1.75, 2.07, 2.40, and 2.87~K~\kms). The red
 contours show emission in the $^{12}$CO(3--2) line integrated over
 52--54.5\,\kms\ (0.31, 0.51, 1.07, 2.10, and
 3.88~K~\kms). The position of V838~Mon is indicated by the star. The green `+'
 symbols mark positions with detected emission in the $^{12}$CO(2--1) line in
 the IRAM single position observations (Off3, Off7, and Off8, as in \citet{kami2}). The `$\times$'
 symbols mark positions with the strongest emission from the
 $^{12}$CO(1--0) line (blue, P10) and $^{12}$CO(3--2) line (red, P32) integrated in
 the 52--54.5\,\kms\ range.}    
\label{echoall}
\end{figure}

The rich observational material presented in Sect.~\ref{obsred} and \citet{kami2} enables us to describe quite accurately the spatial distribution of molecular gas seen in the area of the sky covered by the light echo. The maps made in the CO(1--0) and (3--2) transitions show a compact region of molecular emission tens of arcsec north of V838 Mon. The cloud seen in both transitions is shown in a contour map of Fig.~\ref{echoall}. The observations used to produce the maps are rather limited and show only regions of main emission in CO. The more sensitive single-pointing observations provide an opportunity to constrain even more tightly the gas distribution on the sky. All the positions where point observations led to a detection of the CO(1--0) or CO(2--1) lines are marked in Fig.~\ref{echoall}. We first analyse the single point observations in which the CO(1--0) line detected, which are indicated in Fig.~\ref{echoall} with purple circles. These are the IRAM observations reported in \citet{kami2}. Four of those positions are located outside the region of main molecular emission, which is defined by the contours of the OTF data. The emission found at the position labelled as Off8 has a central radial velocity of $\sim$48~\kms\ and is, most probably, not related to the echo material (see Sect.~\ref{radvel}). The emission lines detected at the three remaining positions outside the contours (Off2, Off4, star position) are weak ($T_{\rm mb}\leq$0.12~K) and reside at a radial velocity of 53.3~\kms, the same as for the main emission region in the OTF data. Although these positions were covered by the OTF observations, the emission lines therein are too weak to be seen in the noisy OTF spectra. The deep IRAM integrations prove however that the region of CO(1--0) emission is more extended than shown by the OTF map. In particular, some weak emission is seen very close to the position of V838 Mon.

The positions with detections of the CO(2--1) line are shown in Fig.~\ref{echoall} with green pluses. The emission at Off8 is not of interest for the same reasons as given above. All the positions with CO(2--1) detected, within the beam-sizes (c.f. Fig.~\ref{iram_rys}), reside within the contours of the CO(1--0) emission. Even though long integrations were obtained at several positions outside the main cloud region, no $J$=2--1 emission was found outside the contours. In particular, there is no emission at the three southern positions where weak lines of CO(1--0) were found (Off2, Off4, position of V838 Mon). Remarkably, the IRAM observations obtained at the position of V838 Mon with an integration time of 6~h gave only an upper limit to the CO(2--1) emission of 3$\sigma$=13.9~mK \kms. This means that the CO(2--1) line is at least 10 times weaker than the corresponding CO(1--0) line at the same position, and 20 times weaker than the CO(1--0) line at the other two neighbouring positions (Off2 and Off4). It may be concluded that the region in the vicinity of V838 Mon lacks CO(2--1) emission.

\subsection{Modelling the physical parameters of the cloud}  

Our observational material provides measurements of multiple transitions at several positions located within the echo region. At least three CO lines were measured for the four positions P10 (5), P32 (5), Off3 (3), and Off7 (3) (where the numbers in brackets indicate the number of measured lines). These measurements can be used to model the physical conditions within the CO cloud. We compare the measured intensities with those simulated for a model cloud with physical conditions varying over a broad range. For radiative transfer calculations and the simulations, we used Radex \citep{radex} (in a version from 2008 Nov. 20). The program and the input data are described in Sect.~\ref{code}. The best-fit set of simulated line intensities was decided using the $\chi^2$ test, where we used the formula     
\begin{equation}
\chi^2=\sum_i^N \left(\frac{I_{{\rm CO,}i}^{\rm obs}-I_{{\rm CO,}i}^{\rm mod}}{\Delta I_{{\rm CO,}i}^{\rm obs}}\right)^2.
\end{equation}
The summation is made for expressions corresponding to each transition measured, where $I_{{\rm CO,}i}^{\rm mod}$  is the simulated intensity, $I_{{\rm CO,}i}^{\rm obs}$ is the measured intensity, and  $\Delta I_{{\rm CO,}i}^{\rm obs}$ is the measurement uncertainty. As the uncertainty, we took the geometric mean of the absolute calibration error and the value of 3$\sigma_{0.1} \Delta V_{\rm ch}$, where $\sigma_{0.1}$ is the rms of the spectrum smoothed to the channel resolution of $\Delta V_{\rm ch}$=0.1~\kms. The absolute calibration errors were assumed to be 25\% for BEARS/NRO, 20\% for HARP/JCMT, 15\% for all the IRAM data at 230 GHz, and 10\% for all the IRAM CO(1--0) data. The assumed BEARS errors are higher than the previously indicated formal calibration errors. This is related to the calibration inconsistency discussed Sect.~\ref{othvsonoff}.

Although in our calculations we included all the positions with multiple CO lines measured, we discuss here only results for positions P10 and P32, where five transitions were observed, and therefore the constraint on physical parameters are strongest. To make the analysis more independent of the uncertain calibration of the BEARS measurements (see Sect.~\ref{othvsonoff}), we compared observations and simulations ({\it i}) for all the measured lines, ({\it ii}) by excluding the BEARS data for $^{12}$CO(1--0), and ({\it iii}) independently of all the BEARS data.

\subsubsection{The code}\label{code}
Radex is a non-LTE radiative transfer code, which uses the
escape probability formulation. We used the program in the
{\it uniform sphere} mode, in which the medium is treated as static,
spherically symmetric and homogenous. 

Input spectroscopic and collisional data required by Radex were taken from the LAMDA database \citep{lamda}, where collisional rates for CO--H$_2$ are taken from \citet{corates}. Though Radex enables computations to be made for many collisional partners, we
considered only collisions with H$_2$. However, we note that in our case atomic hydrogen can also be an important collisional partner of CO,
but that almost nothing is known about the H/H$_2$ ratio in the cloud. We also ignore collisions with He, for which LAMDA does not provide collision rates. The LAMDA database provides CO--H$_2$ collision rates for ortho- and para-H$_2$ molecules. In our
computations, the ortho- to para-H$_2$ ratio was assumed to be thermalized at the given kinetic temperature of the model. As the background radiation field, we implemented only the cosmic microwave background with the black-body temperature of $T_{\rm bg}=2.725$~K. 

The entire radiative transfer in Radex is performed on
rectangular line shapes, i.e. the optical depth is not changed over the
profile. To compare the computed line intensities with the
measured velocity-integrated line intensities, the computed radiation
temperature, $T_R$, is corrected within the program by a factor
$(\sqrt{\pi}/2\sqrt{\ln 2})\times \Delta V = 1.064\,\Delta V$, where $\Delta
V$ is the user-defined Gaussian half-width of the line. The typical observed widths for the $^{12}$CO
lines are $\Delta V = 0.9$~\kms, while for the $^{13}$CO lines $\Delta V = 0.6$~\kms\ (cf. Sect.~\ref{results}). We interpret the broadening of the $^{12}$CO lines as the effect of their moderate or high optical thickness ($\tau \gtrsim 1$). The radiative transfer calculations within Radex were performed with $\Delta V = 0.6$~\kms\ for all lines, but the final integrated intensities were calculated with $\Delta V = 0.9$~\kms\ for the $^{12}$CO lines and $\Delta V = 0.6$~\kms\ for the $^{13}$CO lines.

\subsubsection{The models and results}\label{modelssect}
Using Radex, we generated several grids of models with wide ranges of physical parameters. The initial grid was constructed to place strong constraints on the CO column densities, $N$($^{12}$CO). It was calculated for column densities in the range $10^{13}-10^{21}$~cm$^{-2}$ with values changed in a step of 0.25 dex (decimal exponent). For each value of CO column density, simulations were performed for H$_2$ densities from the set of values given by $n({\rm H}_2)$=[2, 4, 6, 8, 10]$\times10^k$~cm$^{-3}$,  where $k=$2, 3, 4,..., 8, and for kinetic temperatures from the range $T_{\rm kin}$=5--200~K separated by a step of 5~K. For the $^{13}$CO lines, the column density was assumed to be given by the standard ratio $N$($^{13}$CO)=$N$($^{12}$CO)/60 \citep{izoco}. For this grid, the minimum of $\chi^2$ is consistently reached for both positions at $\log[N$($^{12}$CO)/cm$^{-2}$]=16.25 or $\log[N$($^{12}$CO)/cm$^{-2}$]=16.75 when all the BEARS data are excluded. 

As discussed later in Sect.~\ref{dkin}, the analysed molecular gas is most probably located in the outer parts of the Galaxy, where the isotopologue ratio $N$($^{12}$CO)/$N$($^{13}$CO) is known to be higher than the standard value in the solar neighbourhood \cite[e.g.][]{brand}. We therefore generated a grid of models that is similar to the one described above but with $N$($^{13}$CO)=$N$($^{12}$CO)/110, where the high isotopologue ratio was taken after \citet{brand}\footnote{The high $N(^{12}$CO)/$N(^{13}$CO) ratio does not necessarily have to be related to the location of the cloud in the outer Galaxy, but may be related to the isotope-selective nature of the photo-dissociation process of the cloud by the interstellar radiation field \citep[see][]{selective}.}. This grid suggests for both positions that the column density is $\log[N$($^{12}$CO)/cm$^{-2}$]=16.5, but the fit to observations is slightly worse than in the previous grid of models. 

All subsequent simulations were performed with $N$($^{13}$CO)=$N$($^{12}$CO)/60 and $\log[N$($^{12}$CO)/cm$^{-2}$]=16.25, as these values result in best fits to the observations. All the initial models gave $\chi^2$ minima at $T_{\rm kin}$=15$\pm$5\,K.

\begin{figure}\centering
 \includegraphics[angle=270,width=\hsize]{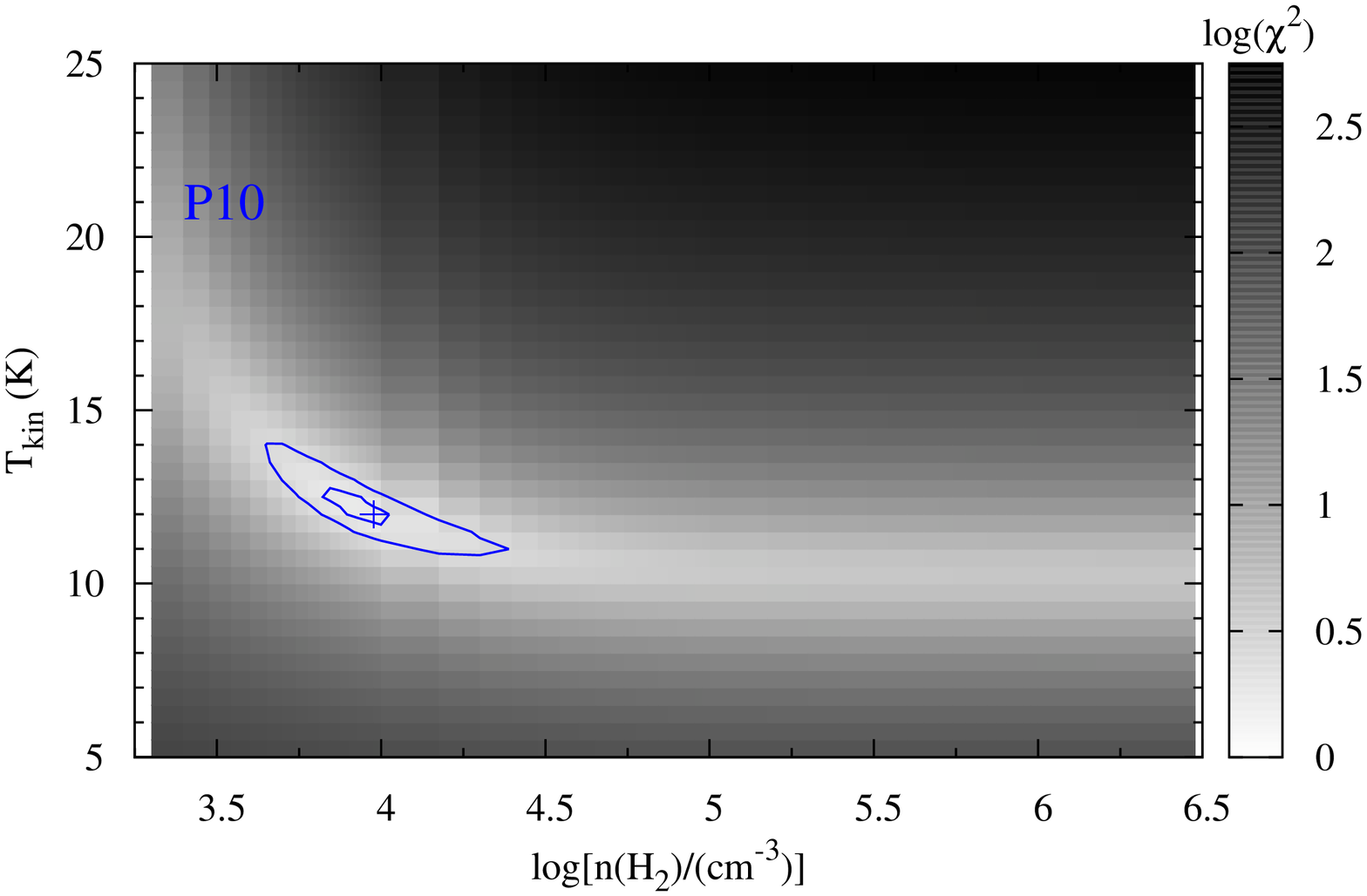}
 \includegraphics[angle=270,width=\hsize]{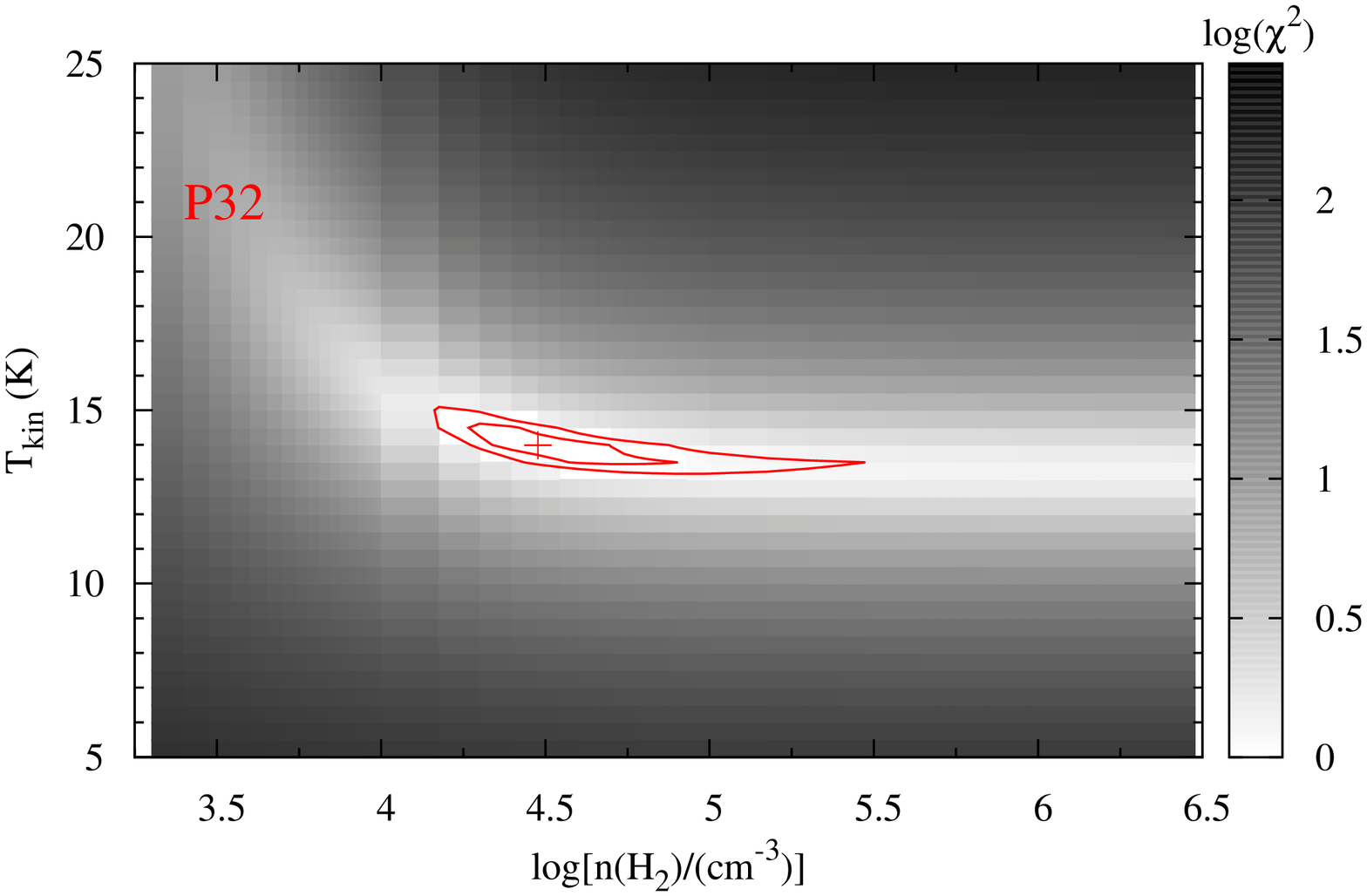}
  \caption{Distributions of $log(\chi^2)$ in the [$T_{\rm  kin}$, $n$(H$_2$)]-plane for the 4 measurements at P10 (top) and P32 (bottom) and $^{12}$CO column density of $\log N$=16.25. The blue contours are drawn for $\chi^2$=1.8 and 2.5, while the red contours mark $\chi^2$=1.0 and 1.2. The pluses indicate the minima.}
\label{chi2}
\end{figure}

With $N$($^{12}$CO) and the CO isotopomer ratio fixed, we generated a denser grid of models at 2~K$\leq T_{\rm kin} \leq$80~K in steps of 0.5~K, and densities given by $n($H$_2)$=$r\cdot10^k$~cm$^{-3}$, where $r$=1,1.5,2,$\dots$,9.5 and $k$=2,3,$\ldots$,7.  The simulated line intensities were compared with the measured intensities excluding the BEARS data for $^{12}$CO(1--0). The distribution of $\chi^2$ in the density--temperature plane is shown in Fig.~\ref{chi2}. 

The minimum of the $\chi^2$ distribution for P10 is at $\chi^2$=1.6 and indicates that  $n($H$_2)$=9.5$\cdot10^3$~cm$^{-3}$ and $T_{\rm kin}$=12~K. For P32, we get a broader but deeper minimum at $\chi^2$=0.9, giving $n($H$_2)$=3$\cdot10^4$~cm$^{-3}$ and $T_{\rm kin}$=14~K. Taking into account uncertainties mainly in the derivations of the column density, the physical conditions towards P10 and P32 may be considered to be identical and represented by the values of $n($H$_2) \approx 10^4$~cm$^{-3}$ and $T_{\rm kin}$$\approx$13~K. In these conditions, the $^{12}$CO lines arising from levels $J$$\leq$3 are of moderate thickness, i.e. 1$\leq \tau \leq$10, and some broadening with respect to the optically thin $^{13}$CO lines is expected. This justifies our modelling procedure described in Sect.~\ref{code}.  

For the derived physical parameters, we also calculated the intensities of the \hcop(1--0) line assuming the molecular abundance of \hcop/H$_2$=2$\cdot$10$^{-9}$ \citep{hcopref}. These predicted line intensities are of the order of 10$^{-5}$\,K, so far below our detection limits. The observed cloud is not dense enough to produce \hcop(1--0)\ line fluxes, which would be easily observed.  

\subsection{Kinematical distance and location in the Galaxy}\label{dkin}
The molecular cloud of interest has a very well defined radial velocity. By fitting a Gaussian to the profile of the emission averaged over the whole cloud, we get a central velocity of $V_{\rm LSR}$=53.34~\kms. The cloud has Galactic coordinates $l$=217\fdg 8 and $b$=1\fdg 0, located close to the outer plane of the Galactic disk. Using rotational curves for this part of the Galaxy, we can estimate the kinematical distance of the cloud. For the rotation curves from \cite{brandblitz} and \cite{fich}, we found that $d$=6.5--7~kpc. Possible streaming motions make this result uncertain by about 1~kpc. For the fully empirical velocity field from \cite{brandblitz}, we get $d$=7.0$\pm$0.3~kpc. The derived values suggest the cloud is located in the Outer Arm (Norma-Cygnus Arm; $d\approx$6.5~kpc) \citep{spiral}. The heliocentric distance of 7~kpc corresponds to the Galactocentric distance of 14~kpc, which indicates that the cloud is located on the outskirts of the Galactic molecular disk. 

\subsection{Column density of H$_2$ and $A_V$}\label{column}

To classify the molecular cloud and reliably constrain its relation to the light echo material,  an estimate of the total column density of H$_2$, or equivalently the reddening $A_V$, is required.  The results of Radex modelling can be used for this purpose. For the considered lines of sight, we obtained  $\log[N$(CO)/cm$^{-2}]$=16.25$\pm$0.25. Assuming the standard CO abundance (with respect to H$_2$) of  $\sim10^{-4}$ \citep{radiobook}, this result can be converted to $\log[N$(H$_2$)/cm$^{-2}]$=20.25$\pm$0.25. The error does not take into account the uncertainty in the CO abundance, which can vary by a factor of about 2.

The total column density of molecules can be derived independently using the X-factor method \citep{xfactor}. For an integrated intensity of the CO(1--0) line, $I_{\rm CO}$, the hydrogen column can be estimated from 
\begin{equation}\label{eee}
N({\rm H}_2)=X_{\rm CO}\cdot \int T_{\rm mb}(^{12}{\rm CO})\,dV \equiv X_{\rm CO} I_{\rm CO}.
\end{equation}
We take $X_{\rm CO}$=2.8 \citep[in units of 10$^{20}$~cm$^{-2}$~K$^{-1}$~km$^{-1}$~s;][]{xfactor}, although values as high as 6 can be found in the literature for the outer Galaxy \citep[see e.g.][]{kami1}. For positions P10 and P32, we get $N$(H$_2$)=$(1.6-1.8) \cdot 10^{21}$~cm$^{-2}$, which should be representative of the densest regions of the cloud. This result is one order of magnitude higher than for the results of Radex modelling, but the actual errors in both estimations are comparable to this difference. For the emission profile averaged over the entire cloud (i.e. within the isophot at 10\% of the peak in the OTF data), we measured $\langle I_{\rm CO}\rangle$=2.88~K~\kms, which gives an ,,average'' H$_2$ column density of $8.1 \cdot 10^{20}$~cm$^{-2}$.  We assume that this value most closely represent the whole molecular cloud. Using the standard conversion formula,  $N$(H)/$E_{B-V}$=5.8$\cdot10^{21}$~cm$^{-2}$~mag$^{-1}$ \citep{bohlin,rach}, we can calculate the corresponding reddening $E_{B-V}$. Ignoring the atomic contribution to the total hydrogen column, i.e. $N$(H)=$N$(H{\small I})$+2N$(H$_2)\approx 2 N$(H$_2)$, we obtain $E_{B-V}$=0.28~mag. For the standard ratio of visual to total extinction of $R_V=$3.1, this corresponds to a visual extinction of $A_V$=0.87~mag. For the column densities at positions P10 and P32 estimated with the X-factor method, we get $A_V$=1.0~mag. This result shows that even in the densest regions of the cloud the extinction is low. This is consistent with the optical HST images of the echo region, which show distant galaxies shining through the cloud (e.g. at $\alpha=07^h04^m05\fs2$, $\delta=-03\degr49\arcmin40\arcsec$).

\subsection{Mass of the cloud}\label{mass}
The mass of the cloud can be estimated using our $^{12}$CO(1--0) data. First, we estimate its virial mass. For a cloud with a spatial density profile of the type $\rho(r)\propto r^{-2}$, with radius $R$ (expressed in pc), and velocity dispersion $\Delta V$ (in \kms), the virial mass (expressed in M$_{\sun}$) is given as \citep{mvir}
\begin{equation}\label{mvireq}
M_{\rm vir}=126\,R\,\Delta V^2.
\end{equation} 
For the 10\% isophot of the CO(1--0) emission and the distance of $d$=6.5~kpc, the radius is $R$=1.5~pc. The emission profile averaged over the entire cloud has a width of $\Delta V$=0.9~\kms. These values give a virial mass of$M_{\rm vir}$=153~M$_{\sun}$.  For a flat density profile, i.e. $\rho(r)$=const, which is less likely, the result would be $M_{\rm vir}$=255~M$_{\sun}$.      

An alternative way of deriving the mass of the cloud is by using the X-factor method. The mass of molecular matter within the cloud can be expressed as 
\begin{equation}\label{mx}
M({\rm H}_2)=X_{\rm CO} \Omega d^2 m_{{\rm H}_2} \langle I_{\rm CO}\rangle,
\end{equation} 
where $\Omega$ is the solid angle of the emission region, $d$ is the distance to the cloud, $m_{{\rm H}_2}$ is the mass of the H$_2$ molecule, and $\langle I_{\rm CO}\rangle$ is the average CO(1--0) line intensity within the cloud. For the solid angle expressed in arcmin$^2$, distance in kpc, line intensity in K~\kms, and $X_{\rm CO}$=2.8$\cdot$10$^{20}$~cm$^{-2}$~K$^{-1}$~km$^{-1}$~s \citep{bohlin}, Eq.~(\ref{mx}) can be rewritten as 
\begin{equation}
M({\rm H}_2)=0.4\Omega d^2 \langle I_{\rm CO}\rangle \, {\rm M}_{\sun}.
\end{equation}
For the solid angle $\Omega$=$\pi r^2$=1.9~arcmin$^2$, the distance of 6.5~kpc, and average intensity of $\langle I_{\rm CO}\rangle$=2.88~K~\kms, we obtain a mass of $M({\rm H}_2)$=92~M$_{\sun}$. Correcting this value for the presence of He, which constitutes about 20\% of the hydrogen mass, we get $M({\rm H}_2+{\rm He})$=111~M$_{\sun}$. Owing to a gradient in the value of  $X_{\rm CO}$ towards the outer Galaxy, the mass may even be twice as large. Taking into account all uncertainties, the derived mass is consistent with the virial masses found above.

\subsection{Classifications of the cloud}
The molecular cloud seen within the light echo region has all the characteristics of an interstellar molecular cloud. In an attempt to classify it into one of the known groups of interstellar clouds, we refer to the classification in  \citet{snow} and \citet{ppIIItab}. 

The densities and column densities (or $A_V$) of molecular matter in the observed cloud are too low to classify it as a dark cloud. The maximal density of our cloud of $\sim$10$^4$~cm$^{-3}$ is in the range of {\it  average} densities typical of dark clouds.
On the other hand, the derived density of the cloud is too high to consider it as a diffuse cloud, for which densities in the range  $n$(H$_2$)=100--500~cm$^{-3}$ are expected. In addition, the derived temperature of $\sim$13~K is too low compared to the range  30--100~K that is typical of diffuse clouds. We therefore classify the cloud as {\it translucent}, which means in-between being dark and diffuse. The derived extinction of $A_V\approx$1.0~mag, which does not take into account a contribution of atomic matter, is consistent with the range $\sim$1--2~mag characterising translucent clouds. The derived mass, dimension, densities, and temperatures also support this classification.

 Observations of molecular clouds show that they have an onion-like structure \citep{snow,cloudstructure}. Translucent clouds are surrounded by a layer of gas with physical parameters typical of diffuse clouds. A layer beyond this may exist, which is dominated by very diffuse atomic gas. It seems that the molecular cloud observed in the field of the light echo has such a complex structure. The CO emission detected close to (positions Off2 and Off4) and at the stellar position, i.e. outside the main region of molecular emission, corresponds to these outer diffuse regions. A  calculations analogous to that described in Sect.~\ref{column} yields $A_V\approx$0.03~mag for this region (the contribution of an atomic component to $A_V$ has not been included).


\section{The CO cloud and the light echo material}\label{connection}
The observed molecular emission centerd at 53.3~\kms\ coincides spatially with the light echo seen in the optical images. This does not necessarily mean that the two environments are physically related, i.e. they do not have to reside at the same distances. Below we argue that the molecular gas and the scattering dust are actually manifestations of the same interstellar cloud located in the vicinity of V838 Mon and its open cluster.

\subsection{Radial velocity} \label{radvel}
The radial velocity of the light-echo material has been determined directly. To our knowledge, there are no spectral observations with a quality and a resolution good enough to obtain these measurements.  There are however some observational findings that enable us to put constrains on the radial velocity. The dust seen in the optical light echo images is certainly located close to V838 Mon, i.e. within several pc \citep{tylecho,sparks}. According to our current understanding of V838 Mon, it is a young object \citep{gromada} and the star is expected to follow the velocity field of its local interstellar environment. If this were the case, the light echo material should have a velocity close to the stellar systemic velocity, i.e.  $V_{\rm LSR}$=54$\pm$1~\kms. This velocity corresponds to the measured central velocity of the SiO maser in V838 Mon \citep{deguchi,atel}, and was confirmed as the velocity of the photosphere of V838 Mon by \citet{uves}. We note that the star is known to be in a binary system with a B3V star, but the separation between the components must be large \citep[$A\,\sim$250\,AU,][]{keckII}, hence the orbital velocity is expected to be small (a few \kms). 
We conclude that the scattering material of the echo has a radial velocity close to the radial velocity of the SiO maser, with an uncertainty of about a few \kms\ related to the orbital motion of V838 Mon.

The radial velocity of the observed CO emission appearing within the echo region is, within the uncertainties, identical to the maser velocity. With respect that written above, this means that the molecular cloud has the same radial velocity as the echo material. A match between radial velocities of objects along the same line of sight is usually taken as strong evidence of the physical association of the objects. We therefore postulate that the molecular cloud seen at $V_{\rm LSR}=53.3$~\kms\ and the dust seen in optical echo images belong to the same interstellar cloud.    

We note that the molecular emission found at $V_{\rm LSR}\sim$48~\kms\ a the radial velocity close to the maser velocity (echo material velocity), but that its spatial distribution (see e.g. Figs. \ref{10maps} and \ref{echoall}) does not indicate any relation with the echo material. This cloud is probably a foreground cloud with respect to the scattering medium around V838 Mon.

\subsection{Distance}
The distance to the scattering cloud of dust seen in the echo images is very well constrained. The most accurate value was found with the polarimetric method of \citet{echopol1}, i.e.  $d$=6.1$\pm$0.6~kpc \citep{sparks}. An independent estimate comes from fitting the main sequence to the photometry of the V838 Mon's cluster members \citep{gromada}, which gives $d$=6.2$\pm$1.2~kpc. These values agree within uncertainties with the kinematical distance derived for the CO cloud of $d$=6.5--7~kpc (see Sect.~\ref{dkin}). This provides further evidence that the dust and gas we observe are components of the same interstellar cloud. 

\subsection{The spatial distribution of molecular gas and dust}
If the dust seen in the light echo and the molecular radio emission originate in the same interstellar cloud, why do they display such different spatial distributions in Fig.~\ref{echoall}? This question is related to the issues of relative distributions of gas and dust in molecular clouds and the complex geometry of the light echo phenomenon.  Both issues are discussed below.

\subsubsection{On the correlation between the spatial distribution of dust and molecular gas}
It is widely known that gas column densities are spatially correlated very well with extinction in molecular clouds. For instance, for dark molecular clouds a very good correlation was found for 2$<$$A_V$$<$30~mag, down to spatial scales of 0.1 pc \citep{alves,lada}. It is often assumed that the correlation continues to the more diffuse clouds with $A_V\leq$2. We note that the correlation of the two distributions is also expressed by the classical relation $N$(H)$\propto E_{B-V}$ \citep{bohlin}. Remarkably, the relation refers to the total hydrogen column density, i.e. molecular and {\it atomic} gas. A molecular cloud is expected to possess an extended 'envelope' of atomic gas of low molecular content and a non-negligible content of dust.

An illustrative example of the relative distribution of dust, atomic matter, and molecular gas comes from an analysis of the interstellar medium within the Pleiades. The cluster is known to reside within a dusty medium clearly seen at ultraviolet and optical wavelengths. In the direction of the young cluster,  CO emission is observed \citep{plejadyCO}, as well as atomic gas traced mainly by \ion{H}{I} emission at 21~cm \citep{plejadyhI}. Comprehensive observations of the cluster and field stars show that the observed dust and gas are indeed located within the cluster \citep{breger1}. The medium has a low density and the derived extinction is in the range  1.0$\lesssim A_V \lesssim 1.6$~mag, which makes it similar to the translucent cloud identified in the echo region.  The relative distribution of atomic gas, molecular matter, and dust within the Pleiades is compared on the web page of S. Gibson\footnote{http://www.naic.edu/$\sim$gibson/pleiades/vla/comparisons.html}. The molecular emission of CO is limited to a very small region and appears close to the highest concentration of dust. The dust is seen in a much larger area of the sky than the CO emission and its distribution correlates very well with the distribution of atomic gas.  Although the interstellar medium in the Pleiades is somewhat extraordinary (the cloud and the cluster do not have a common origin and are thought to be colliding), it may be a good point of reference for what one can expect for the cloud identified close to V838 Mon. The dust can be observed outside the main molecular region, although the highest dust concentration is expected in the main region of molecular emission.  

The optical images of the echo show a bright nebula covering a large region of the sky, which at late epochs extended to 60\arcsec--90\arcsec\  (2--3~pc) from V838 Mon. In particular, the echo is seen in the region of the main molecular emission (see Fig.~\ref{echoall}), but certainly extends beyond it. In light of what was written above, this does not rule out a physical association of the two media and is  expected for a dusty molecular cloud with an extended atomic envelope. The echo images obtained after 2003 \citep{bond2} show a bright echo region close to our position P32, which can be identified with the direction towards the center of the molecular cloud. This bright clump seems to be the center of an extended optical feature, which has the form of a swirl in the late-epoch echo images. The collocation of the center of the feature and the center of our molecular cloud is probably not a coincidence and reflects the physical associations of the scattering dust with the molecular gas. 

\subsubsection{A comparison between the densities of the echo material and the molecular cloud content}

When the distance to a source is known, calibrated images of a light echo in general enable one to reconstruct the dust distribution around the source. For a sequence of images, even a three dimensional distribution of the dusty cloud can be reproduced. Such a project is pending for the light echo of V838 Mon and its result should help us to understand the connection between the dust and the molecular matter. In the present discussion, we estimate the density of matter for only one chosen region of the echo and limit the analysis to one image from 2006 September 10 obtained with ACS/HST in the F814W filter. The local density for a light echo element can be estimated from 
\begin{equation}\label{nH}
n_{\rm H}=\frac{r^2 B_{\rm sca}}{d^2\,\Delta z\,\langle F_{\star}\rangle}\, \left[\int Q_{\rm sca}(\lambda, a)\sigma_{\rm g} \Phi(\theta, \lambda, a)f(a)\,{\rm d}a\right]^{-1},
\end{equation}
where $r$ is the distance of the scattering element from V838 Mon, $B_{\rm sca}$ is the surface brightness of the light echo element, $d$ is the distance to the star, $\Delta z$ is the geometrical thickness of the scattering element along the line of sight (related to the duration of the illuminating flash), $\langle F_{\star} \rangle$ is the flux density of the illuminating source averaged over the entire flash, $Q_{\rm sca}$ is the scattering efficiency; $\sigma_{\rm g}$ is the geometric cross-section of a dust grain, $\Phi$ is a (normalized) phase function for the considered scattering angle, $\theta$, characterizing the echo element, and $f$=$n_{\rm gr}/n_{\rm H}$ is the number density ratio of dust and hydrogen. The integral in Eq.~(\ref{nH}) represents the whole distribution of grain sizes $a$. For our analysis, we chose a bright clump in the echo coincident with position P32. This region is located at an angular  distance of 37\farcs5 from V838 Mon, or equivalently at a projected distance of $\rho$=1.1~pc (at the distance $d$=6~kpc). Surface brightness was measured with a circular aperture of  radius 1\farcs8. The aperture is small enough to avoid contamination by field stars. The surface brightness was measured on the calibrated image (see Sect.~\ref{hst}) in GAIA, giving $B_{\rm sca}({\rm F814W})= 1.5\cdot10^{-21}~{\rm erg~cm}^{-2}~{\rm s}^{-1}~{\rm \AA}^{-1}~{\rm arcsec}^{-2}$. The conversion from counts (electrons) to flux units was done with the scaling factor {\it photlam}=7.03$\cdot 10^{-20}$~erg\,cm$^{-2}$\,\AA\,$(e^-)^{-1}$ \citep{acsfoto}. This measured brightness should be compared to the source intensity in a band of the same throughput as for F814W ($\lambda_{\rm eff}$(F814W)=8332~\AA). Most of the photometric measurements of V838 Mon during the outburst were obtained in the $I_C$ band ($\lambda_{\rm eff}$($I_C$)=7869~\AA). The average brightness of V838 Mon during the main flash was $I_C\approx$5.5~mag, which can be converted\footnote{http://www.stsci.edu/hst/wfpc2/analysis/wfpc2\_cookbook.html} to $m$(F814W)=$I_C+1.21$=6.7~mag. Using the flux-density scale calibrated on Vega \citep[zero point at 25.5~mag,][]{acsfoto}, this gives a flux density, $\langle F_{\star}({\rm F814W})\rangle$, of $1.3 \cdot 10^{-13}~{\rm erg~cm}^{-2}~{\rm s}^{-1}~{\rm \AA}^{-1}$. 

The rapid rise in V838 Mon's brightness in $I_C$ started at the beginning of February 2002 and the object remained very bright in this band for about 80 days \citep[see e.g.][]{vitalyASP}. By solving the echo equation, $z=\rho^2/2ct-ct/2$, for the moments of the beginning and the end of the flash, we get $\Delta z$=0.05~pc. For the moment corresponding to the middle of the flash and using the relation $r=(\rho^2+z^2)^{1/2}$, we get the distance of the scattering element from the star of $r$=1.1~pc. From another echo relation, $\theta=\arcsin(\rho/r)$, we find the scattering angle $\theta$=103\fdg4. The local density of matter can be now expressed as
\begin{eqnarray}
\lefteqn{n_{\rm H}={}}\nonumber\\
&=1.1\!\cdot\! 10^{-22}\!\left[\int\!Q_{\rm sca}(\lambda_{\rm eff}, a)\sigma_{\rm g} \Phi(\theta, \lambda_{\rm eff}, a)f(a)\,{\rm d}a\right]^{-1}\!{\rm cm}^{-1}{\rm sr}^{-1}\label{nd1}&\\
&\equiv\!1.1\!\cdot\! 10^{-22}\,4\pi \big[S(\lambda_{\rm eff})\big]^{-1}\,{\rm cm}^{-3}
=1.3\!\cdot\! 10^{-21}\big[S(\lambda_{\rm eff})\big]^{-1}\!{\rm cm}^{-3}\label{nd2}.
\end{eqnarray}
Values of the integral $S(\lambda, \theta)$ were calculated in \citet{sugerman}, where the author does not use the normalized form of the phase function, hence we make the transformation from Eq.~(\ref{nd1}) to Eq.~(\ref{nd2}). We limit our considerations to silicate dust with a standard ratio of the selective-to-total extinction of $R_V=3.1$ and grain sizes, $a$, in the range 5$\cdot 10^{-4}-1$~$\mu$m \citep[cf.][]{WD01}. \citet{sugerman} presents values for the integral $S$ for three scattering angles,   $\theta$=0\degr, 30\degr, and 180\degr, and we assume his results for $\theta$=180\degr\ to be representative of our case (the phase function changes very slowly between the derived, i.e. $\theta$=103\fdg4, and the adopted angle). For the value of $S$=2$\cdot 10^{-23}$, we then obtain the local density of the echo material towards P32, $n_{\rm H}=65\,{\rm cm}^{-3}$.

The density of the echo material calculated form the optical image can now be compared to the density derived for the same line of sight from radio observations of CO lines. By modelling the line intensities at P32, we found the density of $n_{\rm H}\approx 2n$(H$_2$)=6$\cdot$10$^4$~cm$^{-3}$ (Sect.~\ref{modelssect}). The density of the molecular cloud is almost three orders of magnitude higher than the value derived from the optical image. Thus, it appears that the dust seen in the echo image from 2006 is not in the dense molecular region of the cloud but in its outer (mostly atomic) layers. It seems that at least up until 2006, the inner, molecular parts of the cloud had not been penetrated by the light-echo paraboloids. Taking into account the geometry of a light echo revealing mostly foreground material, the main molecular region should be located ,,behind'' V838 Mon.  
 
\subsubsection{The glow}
An analysis of the HST images of the echo obtained after 2007 in the F606W and F814W filters reveals the presence of a region  close to the position of V838 Mon, which is as bright as the clump that coincides with P32. This bright patch spreads out to $\sim$15\arcsec\ north and 40\arcsec\ south from the star. Hereafter, we call this region the {\it glow}. The flux density of the brightest clumps in this region is 1.3 times higher in F814W than for the clump at P32. At shorter wavelengths, the region at P32 is the brighter one. The red color of the {\it glow} is essential to understanding its origin. After the main eruption, which ended with a $\sim$4.5~mag drop in $I_C$ brightness, V838 Mon has remained fairly bright in the $I_C$ band during its entire subsequent evolution. While in the $V$ band the object faded to about 15~mag, it was still brighter than $\sim$10.5~mag in $I_C$. About half a year after the eruption, a rise in $I_C$ brightness by $\sim$1~mag was observed. We propose that the red {\it glow} observed near V838 Mon does not reflect the light of the main eruption, but is instead related to the significant brightness of the object at red and infrared wavelengths after the main outburst. Scattering of light occurs not in a paraboloidal {\it layer}, but in a {\it filled} paraboloid and the corresponding thickness of the scattering region, $\Delta z$, can be large. This situation is analogous to an echo observed before the end of a flash, as considered e.g. in \citet{tylecho}. 

In the proposed interpretation of the {\it glow}, the scattering occurs at small angles and is therefore very effective, so the emerging nebula can compete in brightness with the echo of the main outburst observed years after the eruption. It is not possible to estimate the local density of the {\it glow} without knowing the dust distribution within the appropriate paraboloid, but it is likely that the effective thickness corresponding to the {\it glow} is much larger than for the region close to P32. Because of the large thickness and high scattering efficiency, the local densities corresponding to the {\it glow} may be much lower than those derived for P32 in the analysis of the optical image. This leads to a conclusion that the region close to P32 coincides with the highest density echo material seen in the optical. This, in turn, shows that the direction toward the center of the molecular cloud coincides with the highest density region of the echo, as expected if the two environments are manifestations of the same interstellar cloud. 

\section{Discussion: the origin of the echo material}\label{disc}
Observations of the ,,infrared echo'' (which we prefer to call the thermal echo) with the Spitzer Space Telescope place initial constraints on the mass of the echoing dust. \citet{spitzer} estimated the mass of the thermally radiating dust to be 1.6~M$_{\sun}$ for the distance of 8~kpc, which when converted to the distance of 6~kpc becomes 0.9~M$_{\sun}$. With the standard gas-to-dust mass ratio of 100, the total mass associated with the echoing cloud is about 90~M$_{\sun}$. The geometry of the thermal echo is analogous to the optical echo, so this mass corresponds to a thin paraboloidal layer within the cloud. The total mass of the cloud may be higher. This estimate is however in excellent agreement with our constraints on the cloud mass of 90--150~M$_{\sun}$ (Sect.~\ref{mass}). Such a massive cloud cannot be the result of mass loss from any single star (or even multiple stars) and is definitely interstellar in nature. Below we investigate its origin in more depth.

V838 Mon is a member of a group of hot stars, which includes three stars of spectral types B3--6 and the B3 companion of V838 Mon \citep{gromada}. All of these stars are seen within the echoing region of the sky but well outside the region of the main molecular emission. The angular distances of those stars from V838 Mon correspond to distances $\lesssim$1.2~pc, while the echo until 2008 was seen at distances reaching at least 3~pc. If the cluster is not extraordinarily elongated in the radial direction, the stars should reside within the interstellar cloud (but not within the molecular part of it). We propose that the dust and gas seen in optical and radio observations is material remaining after the formation of the cluster. 

The cumulative ultraviolet radiation of the group members must have an influence on the interstellar cloud. The cometary shape of the molecular region seems to support this notion. The head of the cometary feature is directed toward the center of the cluster, as often observed for clouds exposed to a strong radiation of hot stars \citep[self-shadowing effect; see e.g.][]{glob1,glob2}. The cloud is therefore in the process of a ,,chemical decay'' \citep{dysocjacja}, which is typical of photon-dominated regions (PDRs). The low column density of our cloud corresponding to $A_V\approx$1~mag, means that it is almost transparent to the interstellar radiation field and the radiation of the cluster.  This implies that it should completely decay on a timescale of Myr. That we observe the cluster to be embedded within the cloud provides strong evidence of its youth. Clusters containing stars of intermediate B sub-types are indeed not observed in association with molecular gas if the cluster is older than 3~Myr \citep[][and references therein]{allen}. Moreover, it is predicted that clusters with no massive stars, as in our case, disperse atomic and molecular gas within 10~Myr \citep{usungaz}. The cluster of V838 Mon and the associated gas should also be in a process of decay in a dynamical sense. The relaxation time of open clusters with small numbers of stars (i.e. less than 36) is estimated to be short, i.e. a few Myr \citep[][and references therein]{allen}. Since the observed cluster is still quite compact ($\sim$1.2~pc), its age should not exceed a few Myr. 	     

It may be concluded that the observed molecular cloud is decaying and the age of the cluster embedded within it is a few Myr, say 3--10~Myr. \citet{gromada} constrained the age of the cluster to be younger than 25~Myr. Our result is consistent with their constraint, but suggests that an even younger age is more probable.  

The above finding is relevant for understanding V838 Mon. Some authors \citep[e.g.][]{bond2} have claimed that the matter seen in the light echo is circumstellar in nature, which would imply  that V838 Mon is an evolved object with a long mass-loss history. This interpretation contradicts our results. The mass of the echo material is too high to be associated with any past mass loss events in V838 Mon. Moreover, as outlined above, the association of V838 Mon with the interstellar cloud suggests that the star must be a young object. 

\begin{acknowledgements} 
We are grateful to K. Menten and M. Walmsley for reading the manuscript and valuable remarks.      
This research was supported by the Polish Ministry of Science and Higher Education under grants N203 004 32/0448 and N203 403939. It also benefited from research
funding from the European Community's sixth Framework Programme under
RadioNet R113CT 2003 5058187. The JCMT data reported in this paper were obtained under project number
{\it M07BI01}. The JCMT is operated by The Joint Astronomy  on
behalf of the Science and Technology Facilities Council of the United Kingdom, the
Netherlands Organisation for Scientific Research, and the National
Research Council of Canada.  Some of the data presented in this paper
were obtained from the Multimission Archive at the Space Telescope
Science Institute (MAST). STScI is operated by the Association of
Universities for Research in Astronomy, Inc., under NASA contract
NAS5-26555. 

\end{acknowledgements}


\begin{thebibliography}{}
\bibitem[Af{\c s}ar \& Bond(2007)]{gromada} 
Af{\c s}ar, M., \& Bond, H.~E.\ 2007, \aj, 133, 387 


\bibitem[Allen et al.(2007)]{allen} 
Allen, L., et al.\ 2007, Protostars and Planets V, 361 

\bibitem[Alves et al.(1999)]{alves} Alves, J., Lada, C.~J., 
\& Lada, E.~A.\ 1999, \apj, 515, 265 

\bibitem[Banerjee et al.(2006)]{spitzer} 
Banerjee, D.~P.~K., Su, K.~Y.~L., Misselt, K.~A., \& Ashok, N.~M.\
2006, \apjl, 644, L57  

\bibitem[Bertoldi \& McKee(1990)]{glob1} 
Bertoldi, F., \& McKee, C.~F.\ 1990, \apj, 354, 529 

\bibitem[Bond(2007)]{bond2}
 Bond, H. E. 2007, ASP Conf. Ser., 363, 130

\bibitem[Bond et al.(2003)]{bond03}
Bond, H. E., Henden, A., Levay, Z. G., et al. 2003, \nat, 422, 405

\bibitem[Bloemen et al.(1986)]{xfactor} 
Bloemen, J.~B.~G.~M., Strong, A. W., Mayer-Hasselwander, H. A., et
   al.\ 1986, \aap, 154, 25  

\bibitem[Brand \& Blitz(1993)]{brandblitz}
Brand J. \& Blitz L. 1993, \aap 275, 67

\bibitem[Brand \& Wouterloot(1995)]{brand} 
Brand, J., \& Wouterloot, J.~G.~A.\ 1995, \aap, 303, 851 


\bibitem[Breger(1987)]{breger1} 
Breger, M.\ 1987, \apj, 319, 754 

\bibitem[Bohlin et al.(1978)]{bohlin} 
Bohlin, R.~C., Savage, B.~D., \& Drake, J.~F.\ 1978, \apj, 224, 132 

\bibitem[Buckle et al.(2009)]{harp1} 
Buckle, J.~V., et al.\ 2009, \mnras, 399, 1026 

\bibitem[Carlqvist(2005)]{spirala} 
Carlqvist, P.\ 2005, \aap, 436, 231 

\bibitem[Corradi \& Munari(2007)]{MunariASP} 
Corradi, R.~L.~M., \& Munari, U.\ 2007, The Nature of V838 Mon and its Light Echo, 363,  

\bibitem[Deguchi et al.(2007)]{deguchi07} 
Deguchi, S., Matsunaga, N., 
\& Fukushi, H.\ 2007, The Nature of V838 Mon and its Light Echo, 363, 81 

\bibitem[Deguchi et al.(2005)]{deguchi} 
Deguchi, S., Matsunaga, N., \& Fukushi, H.\ 2005, \pasj, 57, L25 

\bibitem[Deguchi et al.(2009)]{atel} 
Deguchi, S., Matsunaga, N., Fukushi, H., \& Kami\'nski, T.\ 2009, The Astronomer's Telegram, 1996, 1 

\bibitem[Diaz-Miller et al.(1998)]{dysocjacja} Diaz-Miller, R.~I., 
Franco, J., \& Shore, S.~N.\ 1998, \apj, 501, 192 






\bibitem[Evans et al.(2003)]{evans} Evans, A., Geballe, 
T.~R., Rushton, M.~T., Smalley, B., van Loon, J.~T., Eyres, S.~P.~S., 
\& Tyne, V.~H.\ 2003, \mnras, 343, 1054 

\bibitem[Federman \& Willson(1984)]{plejadyCO} 
Federman, S.~R., \& Willson, R.~F.\ 1984, \apj, 283, 626 

\bibitem[Fich et al.(1989)]{fich}
Fich M., Blitz L., Stark A.A. 1989, \apj 342, 272


\bibitem[Gibson(2007)]{plejadyhI} 
Gibson, S.~J.\ 2007, ASP Conf. Ser., 365, 59 



\bibitem[Goranskij et al.(2007)]{vitalyASP} Goranskij, V.~P., 
Metlova, N.~V., Shugarov, S.~Y., Zharova, A.~V., Barsukova, E.~A., 
\& Kroll, P.\ 2007, The Nature of V838 Mon and its Light Echo, 363, 214 


\bibitem[Heyminck et al.(2006)]{apex} 
Heyminck, S., Kasemann, C., G{\"u}sten, R., de Lange, G., \& Graf,
U.~U.\ 2006, \aap, 454, L21 

\bibitem[Ikeda et al.(2001)]{newstar} 
Ikeda, M., Nishiyama, K., Ohishi, M., \& Tatematsu, K.\ 2001, ASP Conf. Proc., 238, 522 
 

\bibitem[Kami{\'n}ski(2008)]{kami2} 
Kami{\'n}ski, T.\ 2008, \aap, 482, 803

\bibitem[Kami{\'n}ski et al.(2007)]{kami1} 
Kami{\'n}ski, T., Miller, M., \& Tylenda, R.\ 2007, \aap, 475, 569 

\bibitem[Kami{\'n}ski et al.(2009)]{keckI} Kami{\'n}ski, T., 
Schmidt, M., Tylenda, R., Konacki, M., 
\& Gromadzki, M.\ 2009, \apjs, 182, 33 

\bibitem[Kutner \& Ulich(1981)]{chopper}
Kutner, M. L., Ulich, B. L. 1981, \apj, 250, 341

\bibitem[Lada et al.(1999)]{lada} 
Lada, C.~J., Alves, J., \& Lada, E.~A.\ 1999, \apj, 512, 250 

\bibitem[Langer \& Penzias(1993)]{izoco} 
Langer, W.~D., \& Penzias, A.~A.\ 1993, \apj, 408, 539 



\bibitem[Lefloch \& Lazareff(1994)]{glob2} 
Lefloch, B., \& Lazareff, B.\ 1994, \aap, 289, 559 


\bibitem[MacLaren et al.(1988)]{mvir} 
MacLaren, I., Richardson, K.~M., \& Wolfendale, A.~W.\ 1988, \apj, 333, 821 


\bibitem[Mangum et al.(2007)]{otf2} 
Mangum, J.~G., Emerson, D.~T., \& Greisen, E.~W.\ 2007, \aap, 474, 679 

\bibitem[Munari et al.(2002)]{munari} 
Munari, U., et al.\ 2002, \aap, 389, L51 




\bibitem[Palla \& Stahler(2000)]{usungaz} 
Palla, F., \& Stahler, S.~W.\ 2000, \apj, 540, 255 

\bibitem[Rachford et al.(2002)]{rach} 
Rachford, B.~L., et al.\ 2002, \apj, 577, 221 


\bibitem[Rohlfs \& Wilson(2004)]{tools}
Rohlfs, K. \& Wilson, T. L. 2004, {\it Tools of Radio Astronomy}, forth
edition, Springer
 
 \bibitem[Russeil(2003)]{spiral}
 Russeil D. 2003, \aap 397, 133

\bibitem[Sawada et al.(2008)]{otf} 
Sawada, T., et al.\ 2008, \pasj, 60, 445 


\bibitem[Sch{\"o}ier et al.(2005)]{lamda} 
Sch{\"o}ier, F.~L., van der Tak, F.~F.~S., van Dishoeck, E.~F., \&
Black, J.~H.\ 2005, \aap, 432, 369 

\bibitem[Schuster et al.(2004)]{hera} 
Schuster, K.-F., et al.\ 2004, \aap, 423, 1171 

 
\bibitem[Sirianni et al.(2005)]{acsfoto} 
Sirianni, M., et al.\ 2005, \pasp, 117, 1049 

\bibitem[Smith et al.(2008)]{harp2} 
Smith, H., et al.\ 2008, SPIE, 7020,  

\bibitem[Snow \& McCall(2006)]{snow} 
Snow, T.~P., \& McCall, B.~J.\ 2006, \araa, 44, 367 

\bibitem[Soker \& Tylenda(2003)]{soker} 
Soker, N., \& Tylenda, R.\ 2003, \apjl, 582, L105 
  
\bibitem[Sorai et al.(2000)]{sorai} 
Sorai, K., Sunada, K., Okumura, S.~K., Tetsuro, I., Tanaka, A., Natori, K., 
\& Onuki, H.\ 2000, \procspie, 4015, 86 

\bibitem[Sparks(1994)]{echopol1} 
Sparks, W.~B.\ 1994, \apj, 433, 19 


\bibitem[Sparks et al.(2008)]{sparks} 
Sparks, W.~B., et al.\ 2008, \aj, 135, 605 


\bibitem[Sugerman(2003)]{sugerman} 
Sugerman, B.~E.~K.\ 2003, \aj, 126, 1939 

\bibitem[Sunada et al.(2000)]{bears1}
Sunada et al. 2000, SPIE 4015, 237

\bibitem[Thum \& Mauersberger(2007)]{iramnewsletter}
Thum, C. \& Mauersberger, R. 2007, IRAM Newsletter, 68, 10

\bibitem[Turner(2000)]{hcopref} 
Turner, B.~E.\ 2000, \apj, 542, 837 

\bibitem[Tylenda(2004)]{tylecho} 
Tylenda, R.\ 2004, \aap, 414, 223 

\bibitem[Tylenda(2005)]{tyl05} 
Tylenda, R.\ 2005, \aap, 436, 1009 

\bibitem[Tylenda et al.(2009)]{keckII} 
Tylenda, R., Kami{\'n}ski, T., \& Schmidt, M.\ 2009, \aap, 503, 899 

\bibitem[Tylenda et al.(2011)]{uves} 
Tylenda, R., Kami{\'n}ski, T., \& Schmidt, M.\ 2011, in preparation to \aap

\bibitem[Tylenda \& Soker(2006)]{merger} 
Tylenda, R., \& Soker, N.\ 2006, \aap, 451, 223 

\bibitem[Tylenda et al.(2005)]{tylprog}
Tylenda, R., Soker, N., \& Szczerba, R.\ 2005, \aap, 441, 1099 

\bibitem[van der Tak et al.(2007)]{radex} 
van der Tak, F.~F.~S., Black, J.~H., Sch{\"o}ier, F.~L., Jansen,
D.~J., \& van Dishoeck, E.~F.\ 2007, \aap, 468, 627  

\bibitem[van Dishoeck et al.(1993)]{ppIIItab} 
van Dishoeck, E.~F., Blake, G.~A., Draine, B.~T., \& Lunine, J.~I.\ 1993, Protostars and Planets III, 163 

\bibitem[Visser et al.(2009)]{selective} 
Visser, R., van Dishoeck, E.~F., \& Black, J.~H.\ 2009, \aap, 503, 323 


\bibitem[Weingartner \& Draine(2001)]{WD01} 
Weingartner, J.~C., \& Draine, B.~T.\ 2001, \apj, 548, 296 



\bibitem[Wilson et al.(2009)]{radiobook} 
Wilson, T.~L., Rohlfs, K., H\"{u}ttemeister, S.\ 2009, Tools of Radio Astronomy, Springer-Verlag: Berlin  

\bibitem[Wolfire et al.(2010)]{cloudstructure} 
Wolfire, M.~G., Hollenbach, D., \& McKee, C.~F.\ 2010, arXiv:1004.5401, przyjęte do \apj 

\bibitem[Yang et al.(2010)]{corates} 
Yang, B., Stancil, P.~C., 
Balakrishnan, N., \& Forrey, R.~C.\ 2010, \apj, 718, 1062 

\bibitem[Yamaguchi et al.(2000)]{bears2}
Yamaguchi et al. 2000, SPIE 4015, 614

\end{thebibliography}
\end{document}